\documentclass[fleqn,12pt,twoside]{article}
\usepackage[headings]{espcrc1}

\usepackage{multicol}

\RequirePackage{graphicx} 
\newcommand{\D}{\mathrm d} 
\newcommand{\I}{\mathrm i} 

\begin{document} 

\title{Numerical Verification of the Weak Turbulent Model for Swell  Evolution.} 

\author{A.~O.~Korotkevich\address[LandauITP]{Landau Institute for Theoretical Physics RAS,\\
			 2, Kosygin Str., Moscow, 119334, Russian Federation\\}\thanks{{\tt kao@itp.ac.ru}},
        A.~Pushkarev\address[LebedevPI]{Lebedev Physical Institute RAS,\\
			53, Leninsky Prosp., GSP-1 Moscow, 119991, Russian Federation}\address[WSLLC]{Waves and Solitons LLC,\\
			918 W. Windsong Dr., Phoenix, AZ 85045, USA}\thanks{{\tt andrei@cox.net}},
        D.~Resio\address{Coastal and Hydraulics Laboratory,\\
			U.S. Army Engineer Research and Development Center,\\
			Halls Ferry Rd., Vicksburg, MS 39180, USA}
        and
        V.~E.~Zakharov\address[UofA]{Department of Mathematics, University of Arizona,\\
				617 N. Santa Rita Ave., Tucson, AZ 85721, USA}\addressmark[LebedevPI] \addressmark[WSLLC]\addressmark[LandauITP]\thanks{{\tt zakharov@math.arizona.edu}}}
\date{\today} 

\maketitle
\begin{abstract} 
The purpose of this article is numerical verification of the theory of weak turbulence. 
We performed numerical simulation of an ensemble of nonlinearly interacting free 
gravity waves (swell) by two different methods: solution of primordial dynamical equations
describing potential flow of the ideal fluid with a free surface and, solution of the kinetic
Hasselmann equation, describing the wave ensemble in the framework of the theory of weak turbulence.
In both cases 
we observed effects predicted by this theory: frequency downshift, angular 
spreading and formation of Zakharov-Filonenko spectrum $I_{\omega} \sim \omega^{-4}$. To achieve
quantitative coincidence of the results obtained by different methods, one has to 
supply the Hasselmann kinetic equation by an empirical dissipation term $S_{diss}$ modeling 
the coherent effects of white-capping. Using of the standard dissipation terms from operational 
wave predicting model ({\it WAM}) leads to significant improvement on short times,
but not resolve the discrepancy completely, leaving the question about optimal choice of $S_{diss}$ open. 
In a long run {\it WAM} dissipative terms overestimate dissipation essentially. 
\end{abstract} 

\section{Introduction.} The theory of weak turbulence is designed for statistical description of 
weakly-nonlinear wave ensembles in dispersive media. The main tool of weak turbulence theory is 
kinetic equation for squared wave amplitudes, or a system of such equations. Since the 
discovery of the kinetic equation for bosons by Nordheim \cite{Nordheim1928} (see also paper 
by Peierls \cite{Peierls1929}) in the context of solid state physics, this quantum-mechanical 
tool was applied to wide variety of classical problems, including wave turbulence in hydrodynamics, 
plasmas, liquid helium, nonlinear optics, etc. (see monograph by Zakharov, Falkovich and L'vov 
\cite{ZFL1992}). Such kinetic equations have rich families of exact solutions describing 
weak-turbulent Kolmogorov spectra. Also, kinetic equations for waves have self-similar solutions 
describing temporal or spatial evolution of weak -- turbulent spectra. 
 
However, in our opinion, the most remarkable example of weak turbulence is wind-driven sea. The kinetic equation
describing statistically the gravity waves on the surface of ideal liquid was derived by Hasselmann 
\cite{Hasselmann1962}. Since this time the Hasselmann equation is widely used in physical oceanography 
as foundation for development of wave-prediction models such as {\it WAM, SWAN} and {\it WAVEWATCH}: 
it is quite special case between other applications of the theory of weak turbulence due to the 
strength of industrial impact.
 
In spite of tremendous popularity of the Hasselmann equation, its validity and applicability for 
description of real wind-driven sea has never been completely proven. It was criticized by many 
respected authors, not only in the context of oceanography. There are at least two reasons why 
the weak--turbulent theory could fail, or at least be incomplete. 
 
The first reason is presence of the coherent structures. The weak-turbulent theory describes only 
weakly-nonlinear resonant processes. Such processes are spatially extended, they provide weak phase 
and amplitude correlation on the distances significantly exceeding the wave length. However, 
nonlinearity also causes another phenomena, much stronger localized in space. Such 
phenomena -- solitons, quasi-solitons and wave collapses are strongly nonlinear 
and cannot be 
described by the kinetic equations. Meanwhile, they could compete with 
weakly-nonlinear resonant 
processes and make comparable or even dominating contribution in the energy, 
momentum and 
wave-action balance. For gravity waves on the fluid surface the most important 
coherent structures 
are white-cappings (or wave-breakings), responsible for essential 
dissipation of wave energy. 
 
The second reason limiting the applicability of the weak-turbulent theory is 
finite size of any 
real physical system. The kinetic equations are derived only for infinite media, 
where the wave 
vector runs continuous $d$-dimensional Fourier space. Situation is different 
for the wave 
systems with boundaries, e.g. boxes with periodical or reflective boundary 
conditions. The Fourier 
space of such systems is infinite lattice of discrete eigen-modes. If the 
spacing of the lattice is 
not small enough, or the level of Fourier modes is not big enough, the whole 
physics of nonlinear 
interaction becomes completely different from the continuous case. We shall
call effects caused by a finite size of a system ``mesoscopic effects''.
These effects could be important in nature and they certainly should be taken
in to account by anyone who performs numerical simulation of wave turbulence.
 
For these two reasons verification of the weak turbulent theory is an urgent 
problem, important for 
the whole physics of nonlinear waves. The verification can be done by direct 
numerical simulation of 
the primordial dynamical equations describing wave turbulence in nonlinear 
medium. 
 
So far, the numerical experimentalists tried to check some predictions of the 
weak-turbulent theory, such as 
weak-turbulent Kolmogorov spectra. For the gravity wave turbulence the most 
important is 
Zakharov-Filonenko spectrum $F_\omega\sim\omega^{-4}$ \cite{Zakharov1966}. At 
the moment, this spectrum 
was observed in numerous numerical experiments 
\cite{Pushkarev1996}-\cite{Annenkov2006}.
 
The attempts of verification of weak turbulent theory through numerical 
simulation of primordial 
dynamical equations has been started with numerical simulation of 2D optical 
turbulence \cite{DNPZ1992}, 
which demonstrated, in particular, co-existence of weak-turbulent and 
coherent events. 
 
Numerical simulation of 2D-turbulence of capillary waves was done in 
\cite{Pushkarev1996}, \cite{Pushkarev1999}, and \cite{Pushkarev2000}. The main 
results of the simulation consisted in observation of classical regime of weak 
turbulence with spectrum $F_\omega\sim\omega^{-19/4}$, and discovery of non-classical 
regime of ``frozen turbulence'', characterized by absence of energy transfer from 
low to high wave-numbers. The classical regime of turbulence was observed 
on the grid of $256 \times 256$ points at relatively high levels of excitation, 
while the ``frozen'' regime was realized at lower levels of excitation, or more coarse 
grids. The effect of ``frozen'' turbulence is explained by mesoscopic effects:
sparsity of both exact and approximate resonances. The classical regime of turbulence
becomes possible due to broadening of resonances by nonlinearity. The classical and
the frozen regime could coexist. We call this situation ``mesoscopic turbulence''.

In fact, the ``frozen'' turbulence is close to {\it KAM} regime, when the dynamics 
of turbulence is close to the behavior of integrable system \cite{Pushkarev2000}. 

Simulation of the surface gravity waves turbulence for the first time was done simultaneously
by Tanaka~\cite{Tanaka2001} and Onorato at al.~\cite{Onorato2002}. Due to limitations of
calculation performance of that time computers simulations were limited to dynamical
equations and quite short simulation times. After that sime many articles developed
these pioneering results \cite{Dysthe2003}-\cite{Nazarenko2005}. The nonlinear effects
for gravity waves are weaker than for cappilary waves, as a result the influence
of the mesoscopic effects is stronger. It makes the simulations much more difficult.
In the experiments mentioned above, the grid was fine enough to resolve the spectral
tails and observe the weak turbulent Kolmogorov asymptotics $I_k \sim k^{-4}$, but
it was too coarse to avoid mesoscopic effects in the area of spectral peak. The finest
grid $2048\times4096$ was used by Tanaka, however in his experiments the spectral peak
was posed at $k=68$, while the level of nonlinearity was quite small (typical steepness
$\mu \simeq 0.07$). More over, his experiment was very short in terms of characteristic
time of the waves (only several tens periods of the leading wave). If he would continue
his calculations he would observe formation of the weak turbulent Kolmogorov tail as in the
work by Onorato at al.~\cite{Onorato2002}. However his spectral peak was seriously damaged
by mesoscopic effects.

In this article we present the results of simulation of the surface gravity waves turbulence,
namely modelling of swell propagation. All previous experimentalists since Tanaka~\cite{Tanaka2001}
and Onorato at al.~\cite{Onorato2002} performed only one type of experiments --- solution
of the primordial dynamical equations. We made two experiments simultaneously. We solved not
only dynamical equations but also Hasselmann kinetic equation and compared the results.
We think that our results can be considered as the first attempt of direct verification
of Hasselmann kinetic equation.

In the dynamic experiment we used less fine grid than Tanaka ($512\times4096$) but
we posed the spectral peak at $k=300$, and our initial steepness was much higher
than in the Tanaka case ($\mu\simeq0.15$). Due to much higher $k$ and stronger nonlinearity
we managed basically to suppress the mesoscopic effects. The bulk of energy containing modes
satisfies Rayleigh distribution, according to the weak-turbulent scenario. The distribution
has a heavy tail of abnormally intensive harmonics (``oligarchs'' according to terminology,
introduced in the article~\cite{Mesoturb2005}), but they contain no more than 5\% of total
energy.

One important point should be mentioned. In our experiments we observed not only 
weak turbulence, but also additional nonlinear dissipation of the wave energy, 
which could be identified as the dissipation due to white-capping. We observed
fast broadening of the spectra due to multiple harmonics generation. Formation
of the broad spectrum with strong small scale tails can be explained and formation
of quite sharp structures on waves crests. Further development of these tails is
suppressed by an artificial dissipation used in our dynamical experiment.
One can say that in such a way we roughly simulated white-capping phenomenon (continuous
in time dissipation of energy due to multiple acts of wave braking on the very edge of wave crest,
arresting formation of derivative singularity on the sharp wave crest).
Of course we cannot simulate directly wave braking phenomenon but one can say that our method
provide may be rough but reasonable simulation of this effect.

To reach an
agreement with dynamic experiments, we had to add to the kinetic equation
a phenomenological dissipation term $S_{diss}$. In this article we examined
dissipation terms used in the operational wave-prediction models {\it WAM 
Cycle 3} and {\it WAM cycle 4} (hereafter referenced as {\it WAM3} and {\it WAM4} correspondingly). Both of 
these terms overestimate nonlinear dissipation significantly.
Term given in {\it WAM3} gives acceptable results on short periods of time
(time less than $1000T_0$, where $T_0$ is the time period of the leading wave of the initial condition). But at the end of simulation ($t= 3378T_0$) an error in wave action
become close to 30\%. For {\it WAM4} the situation is even worse. If the characteristic wave length
of initial conditions is equal to $22 m$, then $3378T_0$ is a little bit more than only 3 hours of
wave development. Even primitive viscous dissipative term without any simulation of strongly nonlinear
events gives us better results. It means that the question about a reasonable formulae for $S_{diss}$
is open for now.
 
\section{Deterministic and statistic models.} 
In the "dynamical" part of our experiment surface of the fluid was described by two 
functions of horizontal variables 
$x,y$ and time $t$: surface elevation $\eta(x,y,t)$ and velocity potential on 
the surface $\psi(x,y,t)$. They 
satisfy the canonical equations \cite{Zakharov1968} 
\begin{equation} 
\label{Hamiltonian_equations} 
\frac{\partial \eta}{\partial t} = \frac{\delta H}{\delta \psi}, \;\;\;\; 
\frac{\partial \psi}{\partial t} = - \frac{\delta H}{\delta \eta}, 
\end{equation} 
Hamiltonian $H$ is presented by the first three terms in expansion on powers of 
nonlinearity $\nabla \eta$ 
\begin{equation} 
\label{Hamiltonian} 
\begin{array}{l} 
\displaystyle 
H = H_0 + H_1 + H_2 + ...,\\ 
\displaystyle 
H_0 = \frac{1}{2}\int\left( g \eta^2 + \psi \hat k  \psi \right) \D x \D y,\\ 
\displaystyle 
H_1 =  \frac{1}{2}\int\eta\left[ |\nabla \psi|^2 - (\hat k \psi)^2 \right] \D x \D y,\\ 
\displaystyle 
H_2 = \frac{1}{2}\int\eta (\hat k \psi) \left[ \hat k (\eta (\hat k \psi)) + 
\eta\nabla^2\psi \right] \D x \D y. 
\end{array} 
\end{equation} 
Here $\hat k$ is the linear integral operator 
$\hat k =\sqrt{-\nabla^2}$, defined in Fourier space as 
\begin{equation} 
\hat k \psi_{\vec r} = \frac{1}{2\pi} \int |k| \psi_{\vec k} e^{-\I {\vec k} 
{\vec r}} \D \vec k,\; |k|=\sqrt{k_{x}^2 + k_{y}^2}. 
\end{equation} 
Using Hamiltonian (\ref{Hamiltonian}) and equations 
(\ref{Hamiltonian_equations}) one can get the dynamical equations 
\cite{Pushkarev1996}: 
\begin{equation} 
\label{eta_psi_equations} 
\begin{array}{rl} 
\displaystyle 
\dot \eta = &\hat k  \psi - (\nabla (\eta \nabla \psi)) - \hat k  [\eta \hat k  
\psi] +\\ 
\displaystyle 
&+ \hat k (\eta \hat k  [\eta \hat k  \psi]) + \frac{1}{2} \nabla^2 [\eta^2 \hat 
k \psi] +\\ 
\displaystyle 
&\frac{1}{2} \hat k [\eta^2 \nabla^2\psi] + \hat F^{-1}[\gamma_k \eta_k],\\ 
\displaystyle 
\dot \psi = &- g\eta - \frac{1}{2}\left[ (\nabla \psi)^2 - (\hat k \psi)^2 
\right] - \\ 
\displaystyle 
&- [\hat k  \psi] \hat k  [\eta \hat k  \psi] - [\eta \hat k  \psi]\nabla^2\psi  
+ \hat F^{-1}[\gamma_k \psi_k]. 
\end{array} 
\end{equation} 
Here $\hat F^{-1}$ corresponds to inverse Fourier transform. We introduced 
artificial dissipative terms 
$\hat F^{-1}[\gamma_k \eta_k]$ and $\hat F^{-1}[\gamma_k \psi_k]$,
corresponding to pseudo-viscous high frequency damping following recent work \cite{DyachenkoDiasZakharov}.

The model (\ref{Hamiltonian_equations})-(\ref{eta_psi_equations}) was used in 
the numerical experiments \cite{Pushkarev1996} -- \cite{Pushkarev2000}, 
\cite{DKZ2003_Grav}, \cite{DKZ2004}, \cite{Mesoturb2005}, \cite{Nazarenko2005}, \cite{Nazarenko2006}. 
 
Introduction of the complex normal variables $a_{\vec k}$ 
\begin{equation} 
a_{\vec k} = \sqrt \frac{\omega_k}{2k} \eta_{\vec k} + \I \sqrt 
\frac{k}{2\omega_k} \psi_{\vec k}, 
\end{equation} 
where $\omega_k = \sqrt {gk}$, transforms equations 
(\ref{Hamiltonian_equations}) into 
\begin{equation} 
\frac{\partial a_{\vec k}}{\partial t} = -\I\frac{\delta H}{\delta a_{\vec 
k}^{*}}. 
\end{equation} 
 
To proceed with statistical description of the wave ensemble, first, one should 
perform the 
canonical transformation $a_{\vec k} \rightarrow b_{\vec k}$, which excludes the 
cubical terms 
in the Hamiltonian. The details of this transformation can be found in the paper
by Zakharov (1999) \cite{Zakharov1999}.
After the transformation the Hamiltonian takes the form 
\begin{equation} 
\label{Hamiltonian_b_k} 
\begin{array}{c} 
\displaystyle 
H = \int \omega_{\vec k} b_{\vec k} b^{*}_{\vec k}\D\vec k + \frac{1}{4}\int T_{\vec k 
\vec k_1 \vec k_2 \vec k_3} 
b_{\vec k}^{*} b_{\vec k_1}^{*} b_{\vec k_2} b_{\vec k_3}\times\\ 
\displaystyle 
\times\delta_{\vec k + \vec k_1 - \vec k_2 - \vec k_3}\D\vec k_1 \D\vec k_2 \D\vec 
k_3. 
\end{array} 
\end{equation} 
where $T$ is a homogeneous function of the third order: 
\begin{equation} 
T(\varepsilon\vec k, \varepsilon\vec k_1, \varepsilon\vec k_2, \varepsilon\vec 
k_3) = 
\varepsilon^3 T(\vec k, \vec k_1, \vec k_2, \vec k_3). 
\end{equation} 
Connection between $a_{\vec k}$ and $b_{\vec k}$ together with explicit 
expression for 
$T_{\vec k \vec k_1 \vec k_2 \vec k_3}$ can be found, for example, in 
\cite{Zakharov1999}. 
 
Let us introduce the pair correlation function 
\begin{equation} 
<a_{\vec k} a_{\vec k'}^{*}> = g N_{\vec k} \delta(\vec k - \vec k'), 
\end{equation} 
where $N_{\vec k}$ is the spectral density of the wave function. This definition 
of the wave 
action is common in oceanography. 
 
We also introduce the correlation function for transformed normal variables 
\begin{equation} 
<b_{\vec k} b_{\vec k'}^{*}> = g n_{\vec k} \delta(\vec k - \vec k') 
\end{equation} 
Functions $n_{\vec k}$ and $N_{\vec k}$ can be expressed through each other in 
terms of cumbersome 
power series \cite{Zakharov1999} of expansion on $\mu$.
Here $\mu$ is the characteristic steepness defined as follows
\begin{equation}
\label{Steepness_definition}
\mu = \frac{E^2}{N^2}\sqrt{2 E},
\end{equation}
where $E$ is the wave energy and $N$ is the wave action.
Following this definition for the Stokes wave of small amplitude 
\begin{eqnarray*} 
\eta = a\cos(kx),\\ 
\mu \simeq ak. 
\end{eqnarray*}
On deep water their relative difference between $n_{\vec k}$ and $N_{\vec k}$ is of 
the order of $\mu^2$ and can be neglected (in most
cases experimental results shows $\mu \simeq 0.1$).

Spectrum $n_{\vec k}$ satisfies Hasselmann (kinetic) equation \cite{Hasselmann1962} 
\begin{equation} 
\label{Hasselmann_equation} 
\begin{array}{l} 
\displaystyle 
\frac{\partial n_{\vec{k}}}{\partial t}=S_{nl}[n] + S_{diss} + 2\gamma_k n_{\vec 
k}, \\ 
\displaystyle 
S_{nl}[n]=2\pi g^2 \int |T_{\vec{k},\vec{k_1},\vec{k_2},\vec{k_3}}|^2 
\left(n_{\vec{k_1}}n_{\vec{k_2}}n_{\vec{k_3}}+\right.\\ 
\displaystyle 
\left. + n_{\vec{k}}n_{\vec{k_2}}n_{\vec{k_3}} - 
n_{\vec{k}}n_{\vec{k_1}}n_{\vec{k_2}} - 
n_{\vec{k}}n_{\vec{k_1}}n_{\vec{k_3}}\right)\times\\ 
\displaystyle 
\times\delta\left( 
\omega_k+\omega_{k_1}-\omega_{k_2}-\omega_{k_3}\right)\times\\ 
\displaystyle 
\times\delta\left(\vec{k}+\vec{k_1}-\vec{k_2}-\vec{k_3}\right)\,\D 
\vec{k_1}\D\vec{k_2} \D\vec{k_3}. 
\end{array} 
\end{equation} 
Here $S_{diss}$ is an empiric dissipative term, corresponding to white-capping. 
 
Stationary conservative kinetic equation 
\begin{equation} 
S_{nl} = 0 
\end{equation} 
has the rich family of Kolmogorov-type \cite{Kolmogorov1941} exact solutions. 
Among them is Zakharov-Filonenko spectrum \cite{Zakharov1966} for the direct 
cascade of energy 
\begin{equation} 
n_k \sim \frac{1}{k^4}, 
\end{equation} 
and Zakharov-Zaslavsky \cite{ZakharovPhD}, \cite{Zakharov1982} spectra for the 
inverse cascade 
of wave action 
\begin{equation} 
n_k \sim \frac{1}{k^{23/6}}, 
\end{equation} 
 
\section{Deterministic Numerical Experiment.} 
 
\subsection{Problem Setup} 
 
The dynamical equations (\ref{eta_psi_equations}) have been solved in the 
real-space domain $2\pi \times 2\pi$ 
on the grid $512 \times 4096$ with the gravity acceleration set to $g=1$. The 
solution has been performed by the 
spectral code, developed in \cite{KorotkevichPhD} and previously used in 
\cite{DKZ2003_Cap},\cite{DKZ2003_Grav}, 
\cite{DKZ2004},\cite{Mesoturb2005}. We have to stress out that in the current
computations the resolution in 
$Y$-direction (long axis) is better than the resolution in $X$-direction by the factor of 
$8$. 
 
\begin{figure}[htb] 
\centering 
\includegraphics[width=12.5cm]{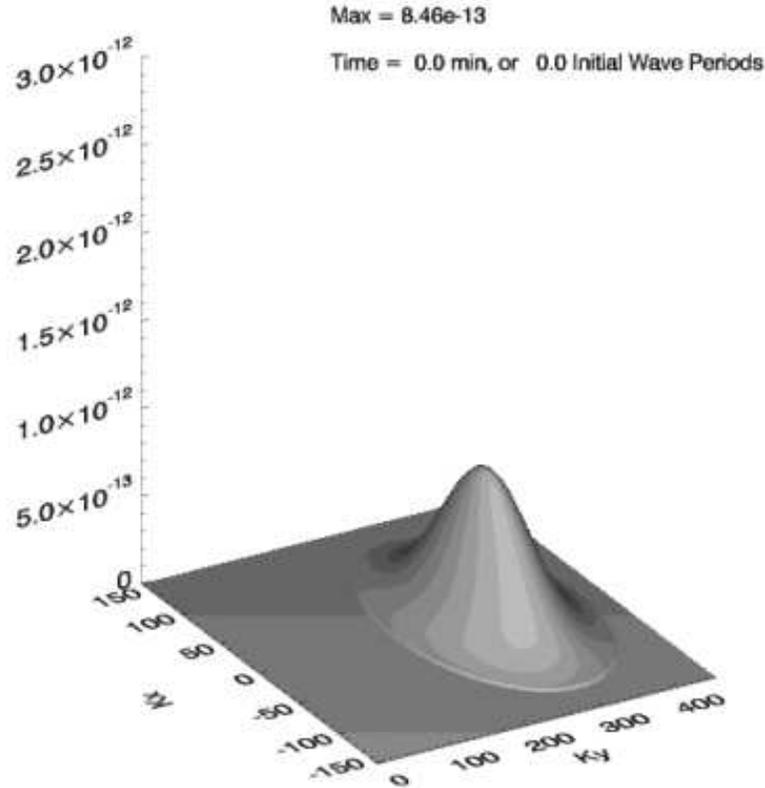} 
\caption{\label{InitialConditions3D}Initial distribution of $|a_{\vec k}|^2$ on 
$\vec k$-plane.} 
\end{figure}

This approach is reasonable if the swell is essentially anisotropic, almost 
one-dimensional. This assumption will be validated by the proper choice of 
the initial data for computation. As the initial condition, we used the  Gaussian-shaped distribution in Fourier 
space (see Fig.~\ref{InitialConditions3D}): 
\begin{equation} 
\label{Dynamic_initial_conditions} 
\begin{array}{l} 
\displaystyle 
\left\{ 
\begin{array}{l} 
\displaystyle 
|a_{\vec k}| = A_i \exp \left(- \frac{1}{2}\frac{\left|\vec k - \vec 
k_0\right|^2}{D_i^2}\right), 
\left|\vec k - \vec k_0\right| \le 2D_i,\\ 
\displaystyle 
|a_{\vec k}| = 10^{-12}, \left|\vec k - \vec k_0\right| > 2D_i, 
\end{array} 
\right.\\ 
\displaystyle 
A_i = 0.92\times10^{-6}, D_i = 60,\\
\displaystyle
\vec k_0 = (0; 300), \omega_0 = \sqrt{g k_0}. 
\end{array} 
\end{equation} 
The initial phases of all harmonics were random. The average steepness of this initial condition was $\mu \simeq 0.15$ (defined in accordance with Eq.\ref{Steepness_definition}).

To realize similar experiment in the laboratory wave tank, one has 
to to generate the waves with wave-length $300$ times less than the length of 
the tank. The width of the tank would not be less than $1/8$  of its length. The 
minimal wave length of the gravitational wave in absence of capillary effects 
can be estimated as $\lambda_{min}\simeq 3 cm$. The leading wavelength should be 
higher by the order of magnitude $\lambda\simeq 30 cm$. 
 
In such big tank of $200 \times 25 $ meters experimentators can observe the 
evolution of the swell until approximately $700 T_0$ -- still less than in our 
experiments. In the tanks of smaller size, the effects of discreetness the 
Fourier space will be dominating, and experimentalists will observe either 
``frozen'', or ``mesoscopic'' wave turbulence, qualitatively different from the 
wave turbulence in the ocean. 

To stabilize high-frequency numerical instability, the damping function has been 
chosen as 
\begin{equation} 
\label{Pseudo_Viscous_Damping} 
\begin{array}{l} 
\displaystyle 
\gamma_k = \left\{ 
\begin{array}{l} 
\displaystyle 
0, k < k_d,\\ 
\displaystyle 
- \gamma (k - k_d)^2, k \ge k_d,\\ 
\end{array} 
\right.\\ 
\displaystyle 
k_d = 1024, \gamma = 5.65 \times 10^{-3}. 
\end{array} 
\end{equation} 
 
The simulation was performed until $t = 1225$, which is equivalent to $3378\,T_0$, 
where $T_0=2\pi/\sqrt{k_0}$ is the 
period of the wave, corresponding to the maximum of the initial spectral 
distribution. 
 
\subsection{Zakharov-Filonenko spectra} 
 
\begin{figure}[htb] 
\centering 
\includegraphics[width=12.5cm,angle=0]{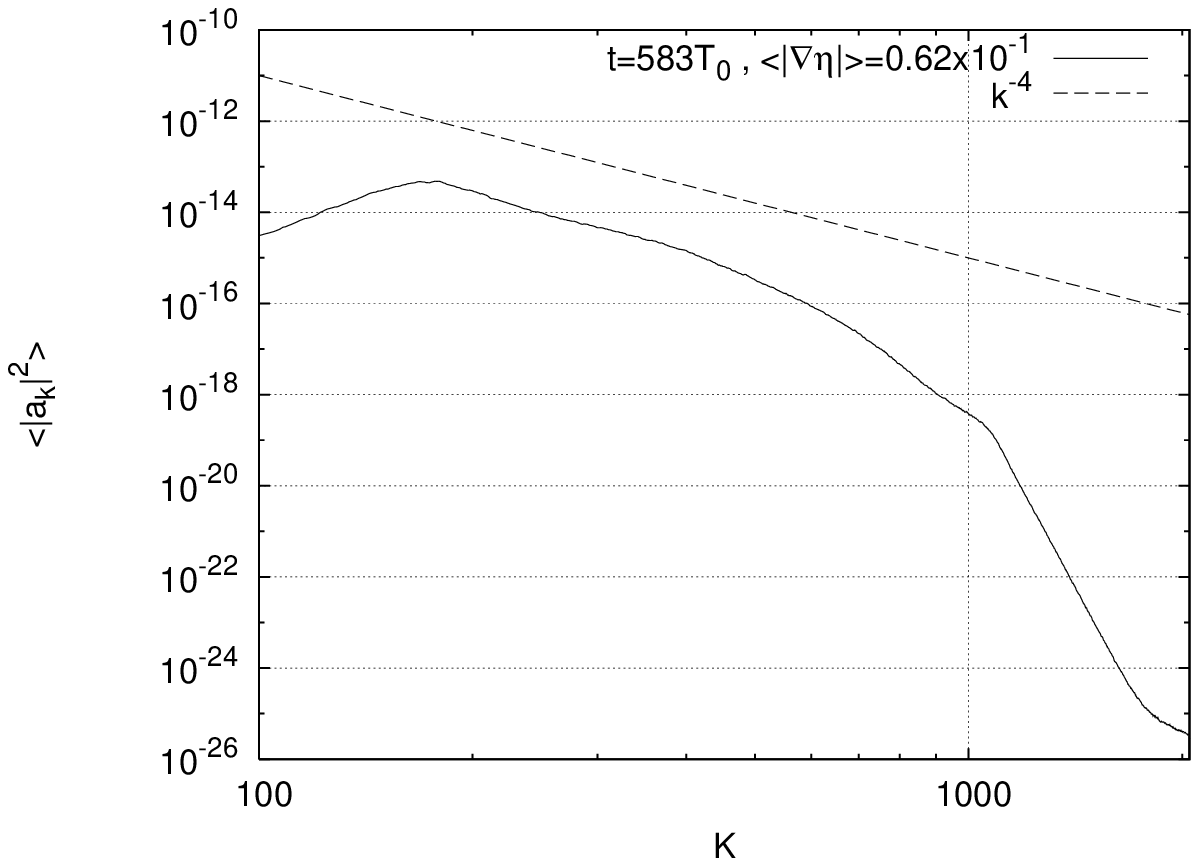} 
\caption{\label{Kolmogorov_k}Angle-averaged spectrum $n_k=<|a_{\vec k}|^2>$ in a 
double logarithmic scale. The tail of distribution 
fits to Zakharov-Filonenko spectrum.} 
\end{figure}

Like in the previous papers 
\cite{Onorato2002},\cite{DKZ2003_Grav},\cite{DKZ2004} and \cite{Mesoturb2005}, 
we observed fast formation of the spectral tail, described by Zakharov-Filonenko 
law for the direct cascade 
$n_k \sim k^{-4}$ \cite{Zakharov1966} (see Fig.\ref{Kolmogorov_k}). In the 
agreement with \cite{Mesoturb2005}, 
the spectral maximum slowly down-shifts to the large scales region, which 
corresponds to the inverse cascade 
\cite{ZakharovPhD},\cite{Zakharov1982}. 
 
\begin{figure}[htb] 
\centering 
\includegraphics[width=12.5cm,angle=0]{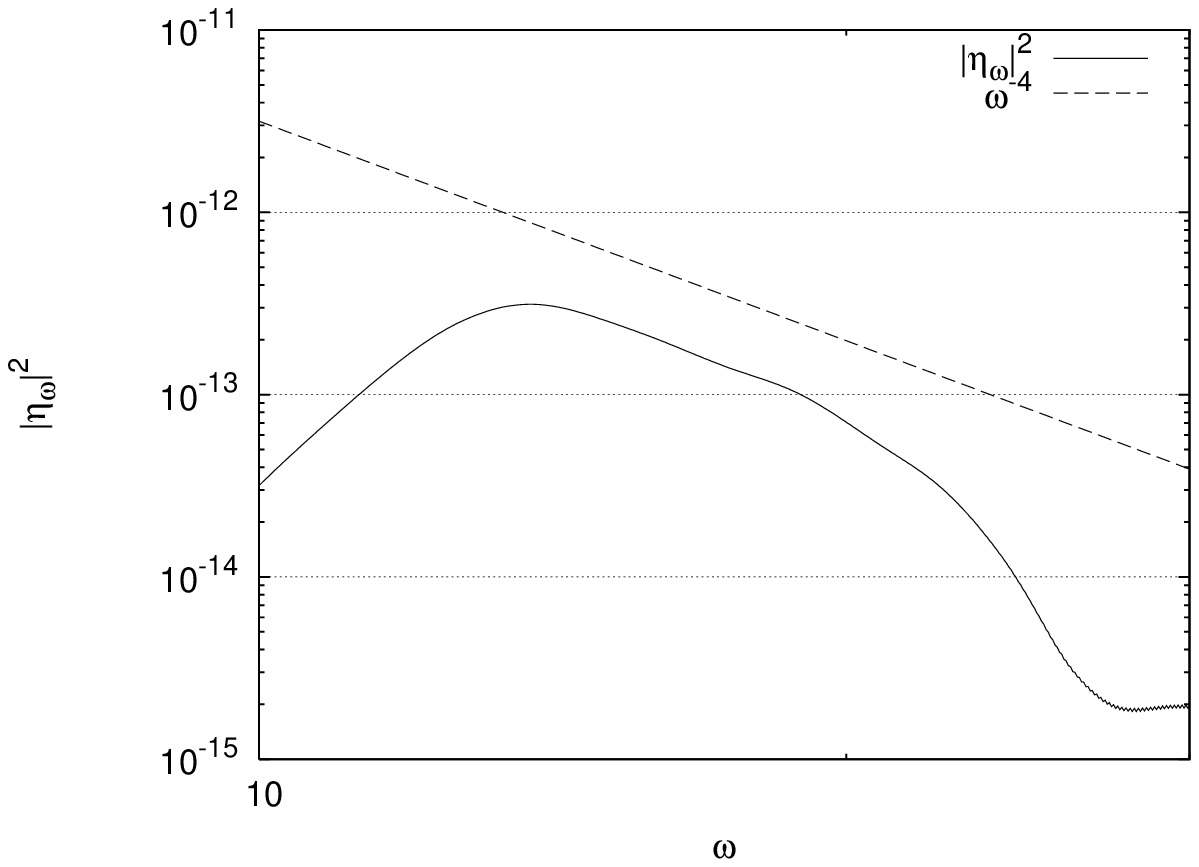} 
\caption{\label{Kolmogorov_omega}Energy spectrum in a double logarithmic scale. 
The tail of distribution fits to asymptotic $\omega^{-4}$.} 
\end{figure}

Also, the direct measurement of energy spectrum has been performed during the 
final stage of the simulation, 
when the spectral down shift was slow enough.  This experiment can be 
interpreted as the ocean buoy record -- 
the time series of the surface elevations has been recorded at one point of the 
surface during 
$T_{buoy} \simeq 300 T_0$. The Fourier transform of the autocorrelation function 
\begin{equation} 
E(\omega) = \frac{1}{2\pi}\int\limits_{-T_{buoy}/2}^{T_{buoy}/2}<\eta(t+\tau)\eta(\tau)>e^{i\omega t}\D \tau \D t. 
\end{equation} 
allows to detect the direct cascade spectrum tail proportional to $\omega^{-4}$ 
(see Fig.\ref{Kolmogorov_omega}), 
well known from experimental observations 
\cite{Toba1973},\cite{Donelan1985},\cite{Hwang2000}. 

In this paper we had no intention to improve results of the previous papers. Inertial interval of the
angle-averaged spectra ($\omega$-spectrum also is angle-averaged because it depends only on frequency
$\omega \sim \sqrt{k}$) is limited due to anisotropy of the integration domain. In this case we have less than
one decade interval and we cannot find an exponent of the Kolmogorov tail surely, however
this work was already done in several previous articles (see for instance~\cite{DKZ2004}). Here we limit
our observations by qualitative correspondance.
 
\subsection{Is the weak-turbulent scenario realized?} 
 
Presence of Kolmogorov asymptotic in spectral tails, however, is not enough to validate 
applicability of the weak-turbulent scenario for description of wave ensemble. We have also be sure 
that statistical properties of the ensemble correspond to weak-turbulent theory assumptions. 
 
\begin{figure}[htb]
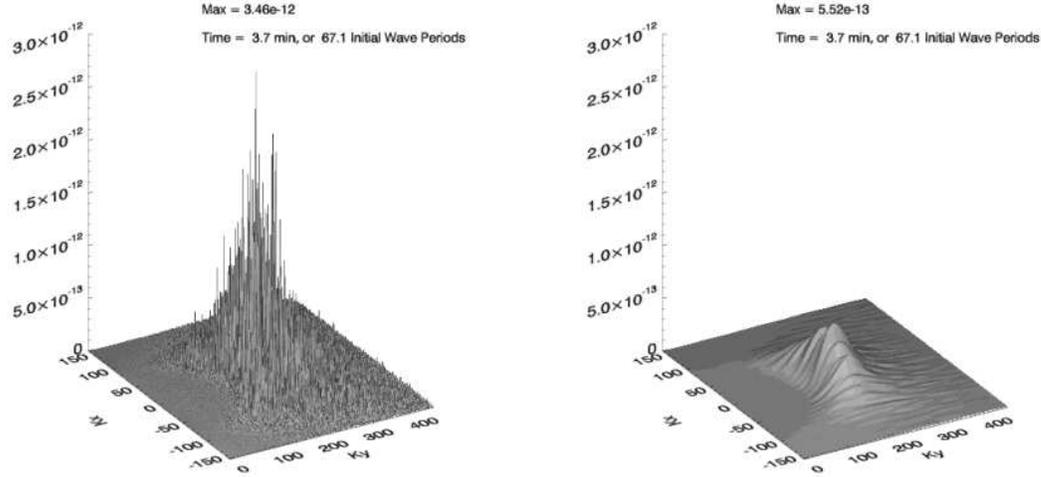
 
\centering 
\includegraphics[width=7.5cm]{figures/bin.cut.a_k_c_3D.072000.eps2}
\includegraphics[width=7.5cm]{figures/bin.cutdomain.filtered.eps2}  
\caption{\label{RawDistribution}\label{SmoothedDistribution}Surface $|a_{\vec k}|^2$ before (left) and after
(right) low-pass filter at the moment of time $t\simeq67T_0$.}
\end{figure}

One should stress out that at the beginning of our experiments $|a_{\vec k}|^2$ is a smooth function
of $\vec k$. Only the phases of the individual waves are random. As shows numerical simulation, the 
initial condition (\ref{Dynamic_initial_conditions}) (see Fig.\ref{InitialConditions3D}) 
does not preserve its smoothness -- it becomes rough virtually immediately (see Fig.\ref{RawDistribution}). 
The picture of this roughness is remarkably preserved in many details, even as 
the spectrum down-shifts as a whole. This roughness does not contradict the 
weak-turbulent theory. According to this theory, the wave ensemble is almost 
Gaussian, and both real and imaginary parts of each separate harmonics are 
not-correlated. However, also according to the weak-turbulent theory, the spectra 
must become smooth after averaging over long enough time of more than $1/\mu^2$ 
periods.  Earlier we observed such restoration of the smoothness in the numerical 
experiments of the $MMT$ model (see \cite{ZGDP},\cite{DPZ}, \cite{DGPZ} and \cite{GZD}). However, 
in the experiments discussed in the article, the roughness still persists and the 
averaging does not suppresses it completely. It can be explained by sparsity of 
the resonances. 
 
Resonant conditions are defined by the system of equations: 
\begin{equation} 
\label{resonant_conditions} 
\begin{array}{c} 
\displaystyle 
\omega_k + \omega_{k_1} = \omega_{k_2} + \omega_{k_3},\\ 
\displaystyle 
\vec k + \vec k_1 = \vec k_2 + \vec k_3, 
\end{array} 
\end{equation} 
These resonant conditions define five-dimensional hyper-surface in 
six-dimensional space $\vec k, \vec k_1, \vec k_2$. In a finite system, 
(\ref{resonant_conditions}) turns into Diophantine equation. Some solutions of 
this equation are known \cite{DZ1994}, 
\cite{Nazarenko2005}. In reality, however, the energy transport is realized not 
by exact, but approximate resonances,
posed in a layer near the resonant surface and defined by 
\begin{equation} 
\label{resonant_layer} 
|\omega_{k} + \omega_{k_1} - \omega_{k_2} - \omega_{k + k_1 - k_2}| \le 
\Gamma, 
\end{equation} 
where $\Gamma$ is a characteristic inverse time of nonlinear interaction. 
 
In the finite systems $\vec k, \vec k_1, \vec k_2$ take values on the nodes of 
the discrete grid. The weak turbulent approach is valid, 
if the density of discrete approximate resonances inside the layer 
(\ref{resonant_layer}) is high enough. In our case the lattice 
constant is $\Delta k = 1$, and typical relative deviation from the resonance 
surface 
\begin{equation} 
\frac{\Delta \omega}{\omega} \simeq \frac{\omega_k'}{\omega} \Delta k = 
\frac{\omega_k'}{\omega} \simeq \frac{1}{600} 
\simeq 2\times10^{-3}. 
\end{equation} 
Inverse time of the interaction $\Gamma$ can be estimated from our numerical 
experiments: wave amplitudes change essentially during 30 
periods, and one can assume: $\Gamma/\omega \simeq 10^{-2} \gg \frac{\delta 
\omega}{\omega}$. It means that the condition for the 
applicability of weak turbulent theory is typically satisfied, but the reserve 
for their validity is rather modest. As a result, 
some particular harmonics, posed in certain ``privileged'' point of $k$-plane 
could form a network of almost resonant quadruplets 
and realize significant part of energy transport. Amplitudes of these harmonics 
exceed the average level essentially. This effect 
was described in the article \cite{Mesoturb2005}, where such ``selected few'' 
harmonics were called ``oligarchs''. If oligarchs 
realize most part of the energy flux, the turbulence is mesoscopic, not weak. 
 
\subsection{Statistics of the harmonics} 
 
According to the weak-turbulent scenario, statistics of the $a_{\vec k}(t)$ in 
any given $\vec{k}$ should be close to 
Gaussian. It presumes that the $PDF$ for the squared amplitudes is 
\begin{equation}
\label{GaussPDF}
P(|a_{\vec k}|^2) \simeq \frac{1}{D}e^{-|a_{\vec k}|^2/D}, 
\end{equation} 
where $D = <|a_{\vec k}|^2>$ is the mean square amplitude.

To check the equation (\ref{GaussPDF}), we need to find the way for calculation of $D(\vec k)$.
If the ensemble is stationary in time, $D(\vec k)$ could be found for any given $\vec{k}$ by the time averaging.
In our case, the process is non-stationary, and we have a problem with determination of $D(\vec k)$. 
 
To resolve this problem, we used low-pass filtering instead of time averaging. The low-pass filter was taken in the form 
\begin{equation} 
\begin{array}{l} 
\displaystyle 
f(\vec n) = e^{-(|\vec n|/\Delta)^3}, \\ 
\Delta=0.25 \cdot N_x/2,\,\, N_x=4096. 
\end{array} 
\end{equation} 
where $\vec{n}$ is integer vector running $K$-space.

This choice of the low-pass filter  preserves the  values of total energy, wave action and the total momentum within 
three percent accuracy. The shape of low-pass filtered function is presented on Fig.\ref{SmoothedDistribution}. 

\begin{figure}[htb] 
\centering 
\includegraphics[width=12.5cm]{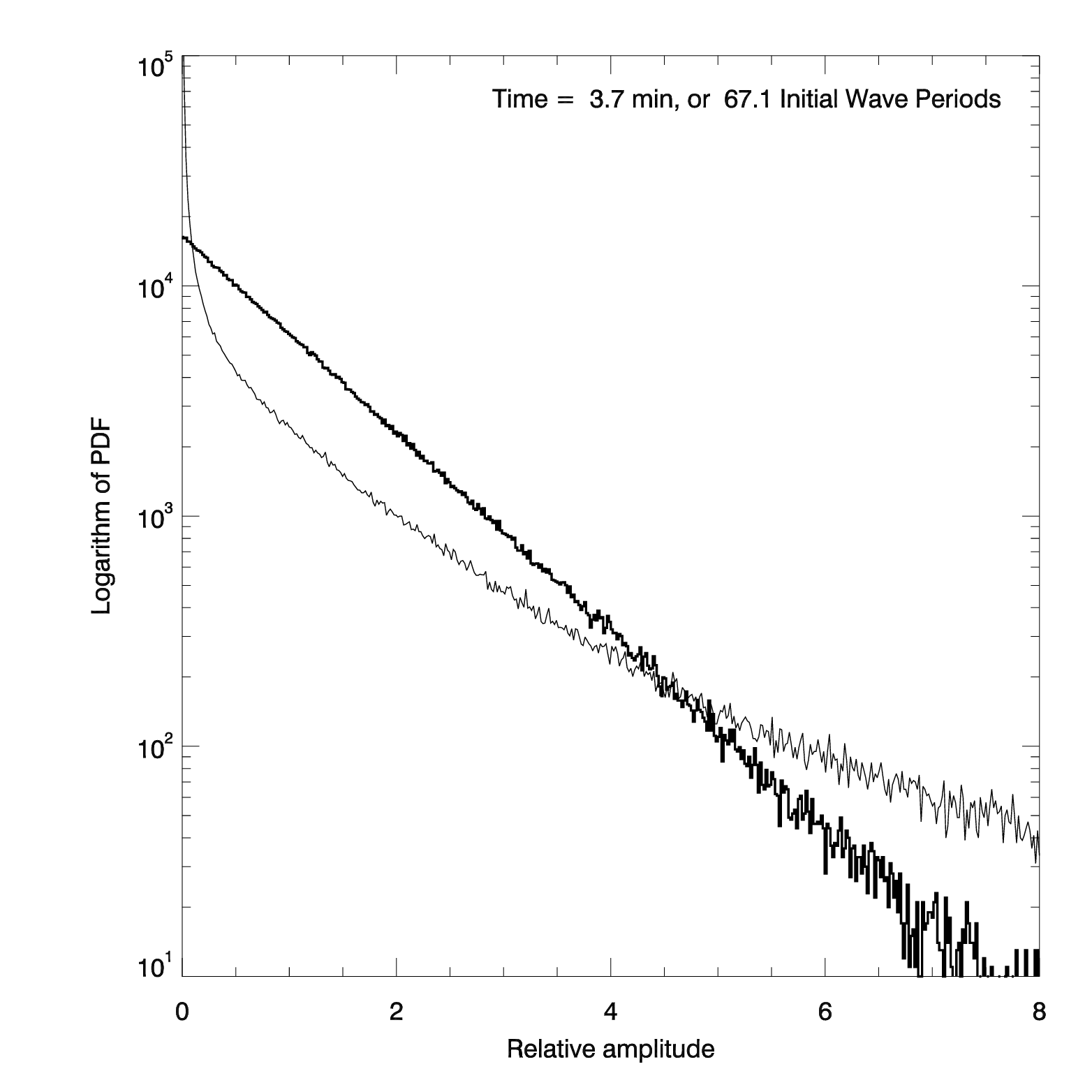} 
\caption{\label{DistributionFunction70}Probability distribution function
for relative squared amplitudes $|a_k|^2/<|a_k|^2>$. $t\simeq67T_0$. Thin and bold lines -- averaging over dissipation and dissipation-free regions of $\vec{K}$-space correspondingly.}
\end{figure}

\begin{figure}[htb] 
\centering 
\includegraphics[width=12.5cm]{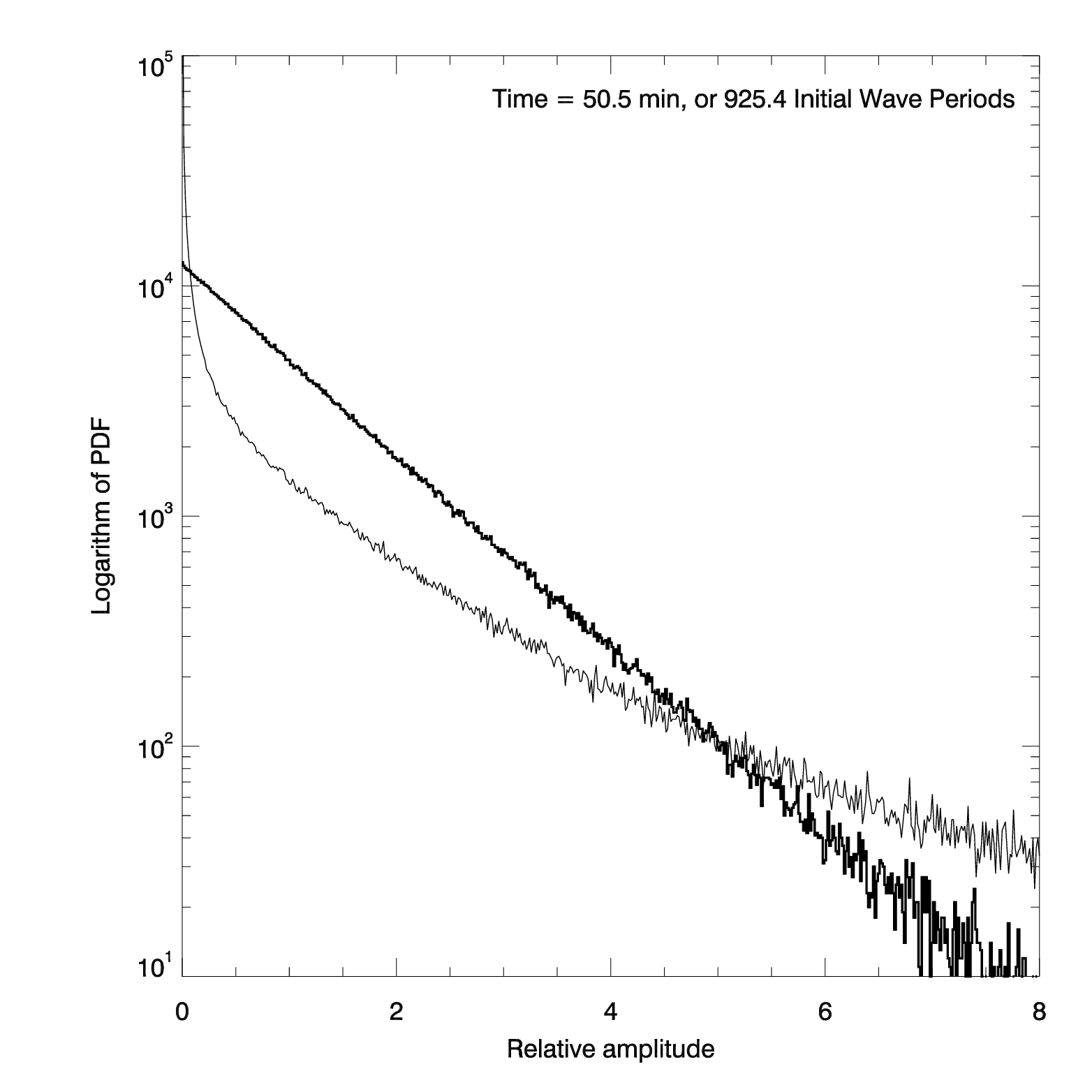} 
\caption{\label{DistributionFunction933}Probability distribution function 
for relative squared amplitudes $|a_k|^2/<|a_k|^2>$. $t\simeq925T_0$. Thin and bold lines -- averaging over dissipation and dissipation-free regions of $\vec{K}$-space correspondingly.} 
\end{figure}

Now it is possible to average the $PDF$ over different areas in $k$-space. The 
results for two different moments of time $t\simeq 67.1 T_0$ and $t\simeq925T_0$ 
are presented in Fig.\ref{DistributionFunction70} and 
Fig.\ref{DistributionFunction933}. 
The thin line gives $PDF$ after averaging over dissipation region harmonics, while bold 
line presents averaging over the non-dissipative area $|\vec k| < k_d = 1024$. One can 
see that statistics in the last case is quite close to the Gaussian, while in the 
dissipation region it deviates from Gaussian distribution. However, deviation from the 
Guassianity in the dissipation region does not create any problems, since the ``dissipative'' 
harmonics do not contain any essential amount of the total energy, wave action and momentum. 
 
One should remember, that the bold lines in the Fig.\ref{DistributionFunction70} and 
Fig.\ref{DistributionFunction933} are the results of averaging over a million of harmonics.
Among them there is a population of ``selected few'', or ``oligarchs'', with the amplitudes 
exceeding the average value by the factor of more than ten times. The oligarchs exist 
because our grid is still not fine enough. 
The contribution of such oligarchs in our case to the total wave action does not 
exceed 4\%. 
Fifteen leading oligarchs at some moment of time are presented in the Appendix \ref{APP:ForbesList}. 
 
\clearpage
\subsection{Two-stage evolution of the swell} 
 
\begin{figure}[ht]
\centering
\includegraphics[scale=0.55]{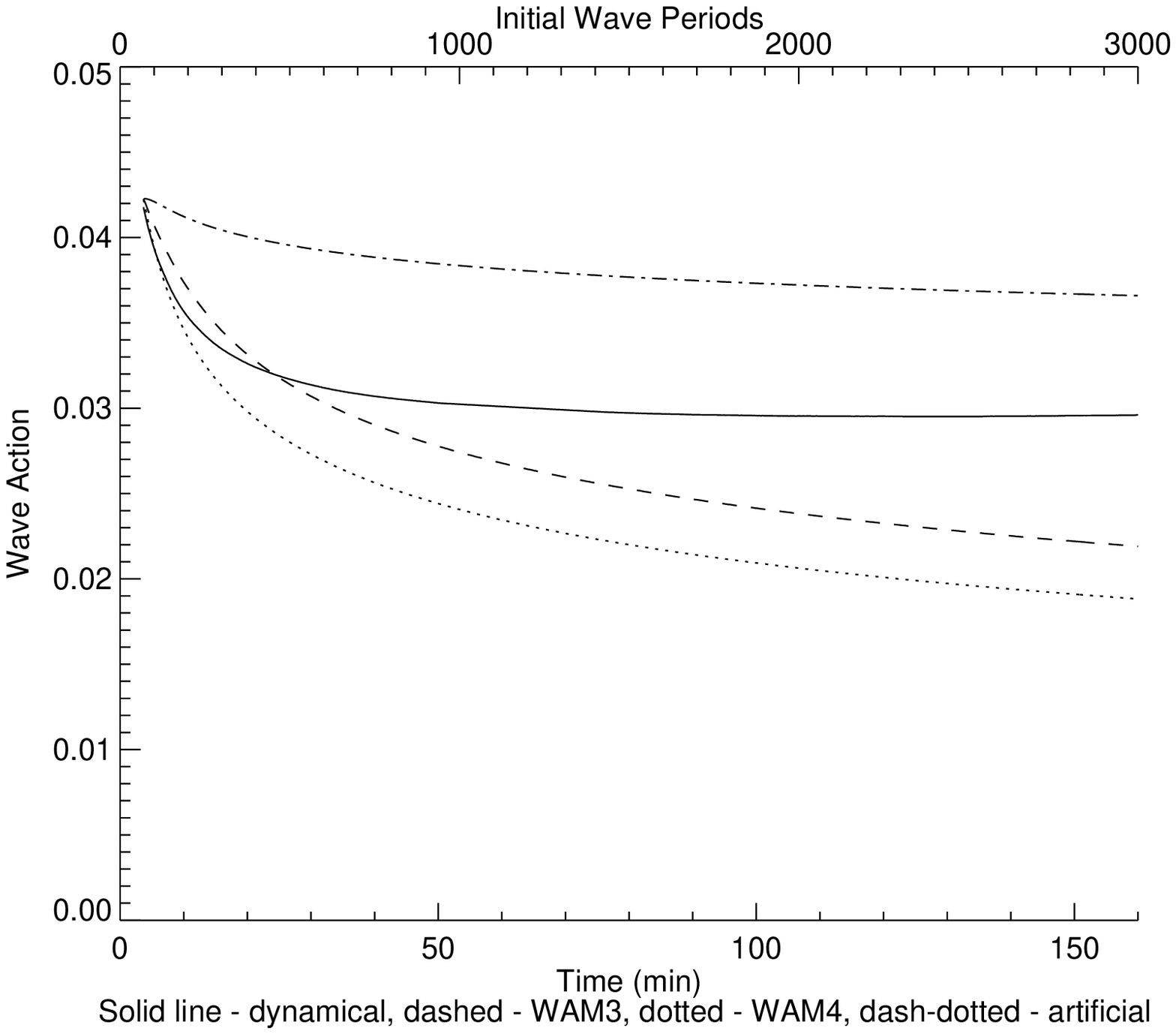}
\caption{Total wave action as a function of time. Solid line --- dynamical equations,
dashed-dotted line --- kinetic equation with artificial viscosity,
dashed line --- kinetic equation with {\it WAM3} damping term,
dotted line --- kinetic equation with {\it WAM4} damping term}
\label{Action} 
\end{figure}

\begin{figure}[ht] 
\centering
\includegraphics[scale=0.55]{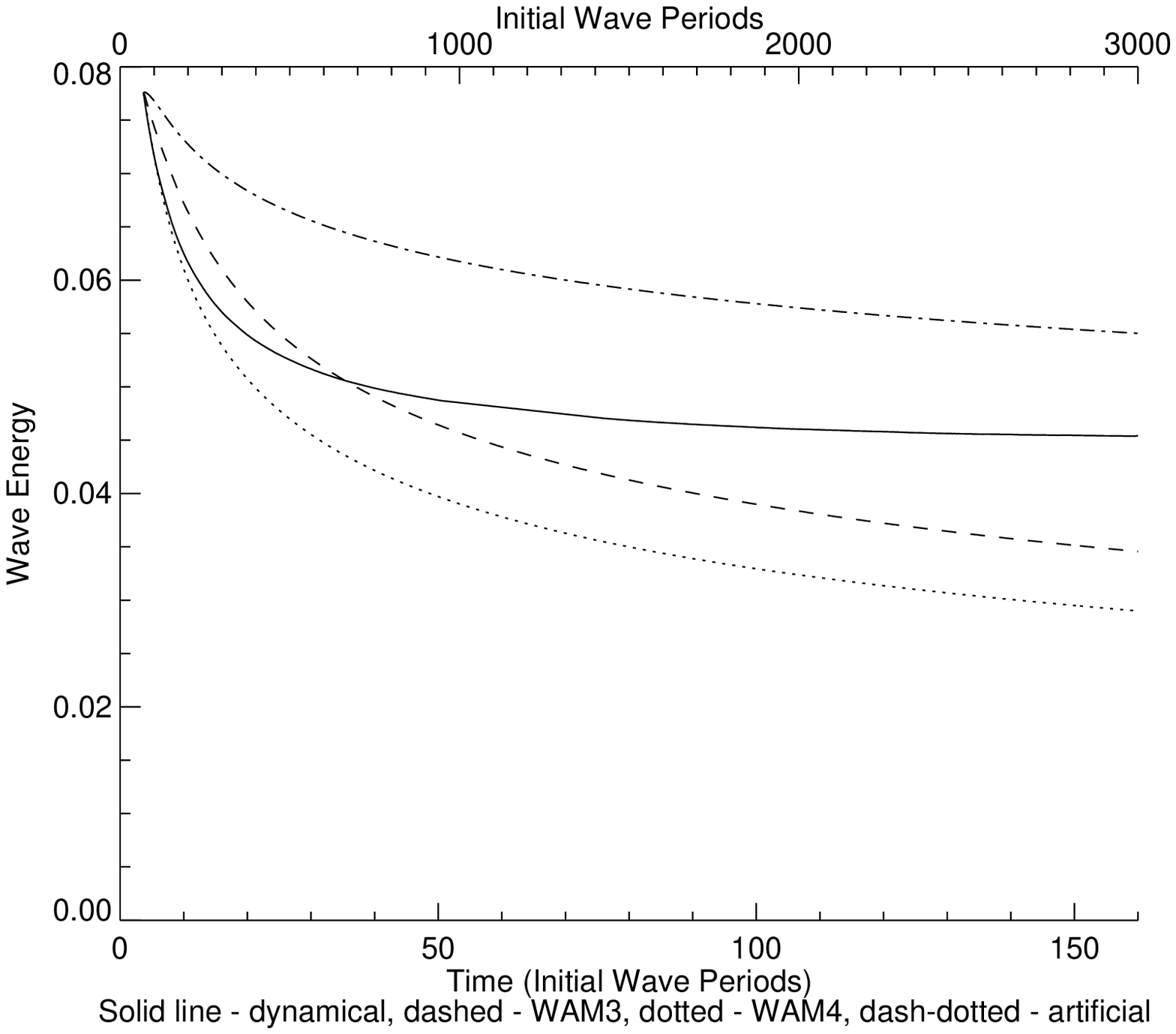} 
\caption{Total wave energy as a function of time. Solid line --- dynamical equations,
dashed-dotted line --- kinetic equation with artificial viscosity,
dashed line --- kinetic equation with {\it WAM3} damping term,
dotted line --- kinetic equation with {\it WAM4} damping term}
\label{Energy} 
\end{figure}

\begin{figure}[ht] 
\centering
\includegraphics[scale=0.55]{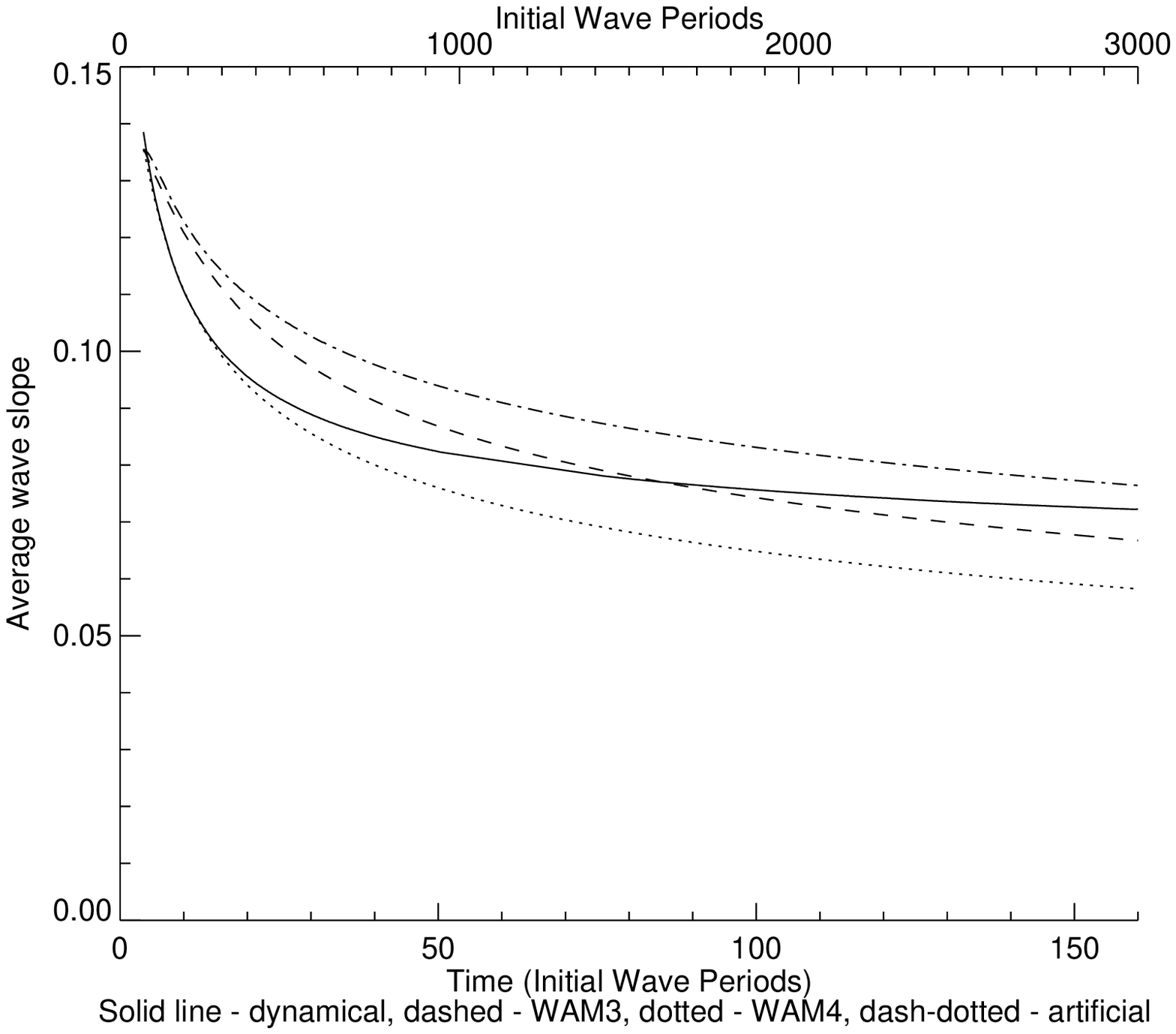} 
\caption{Average wave slope as a function of time. Solid line --- dynamical equations,
dashed-dotted line --- kinetic equation with artificial viscosity,
dashed line --- kinetic equation with {\it WAM3} damping term,
dotted line --- kinetic equation with {\it WAM4} damping term}\label{Slope_ArtVisc} 
\end{figure}

\begin{figure}[ht] 
\centering
\includegraphics[scale=0.55]{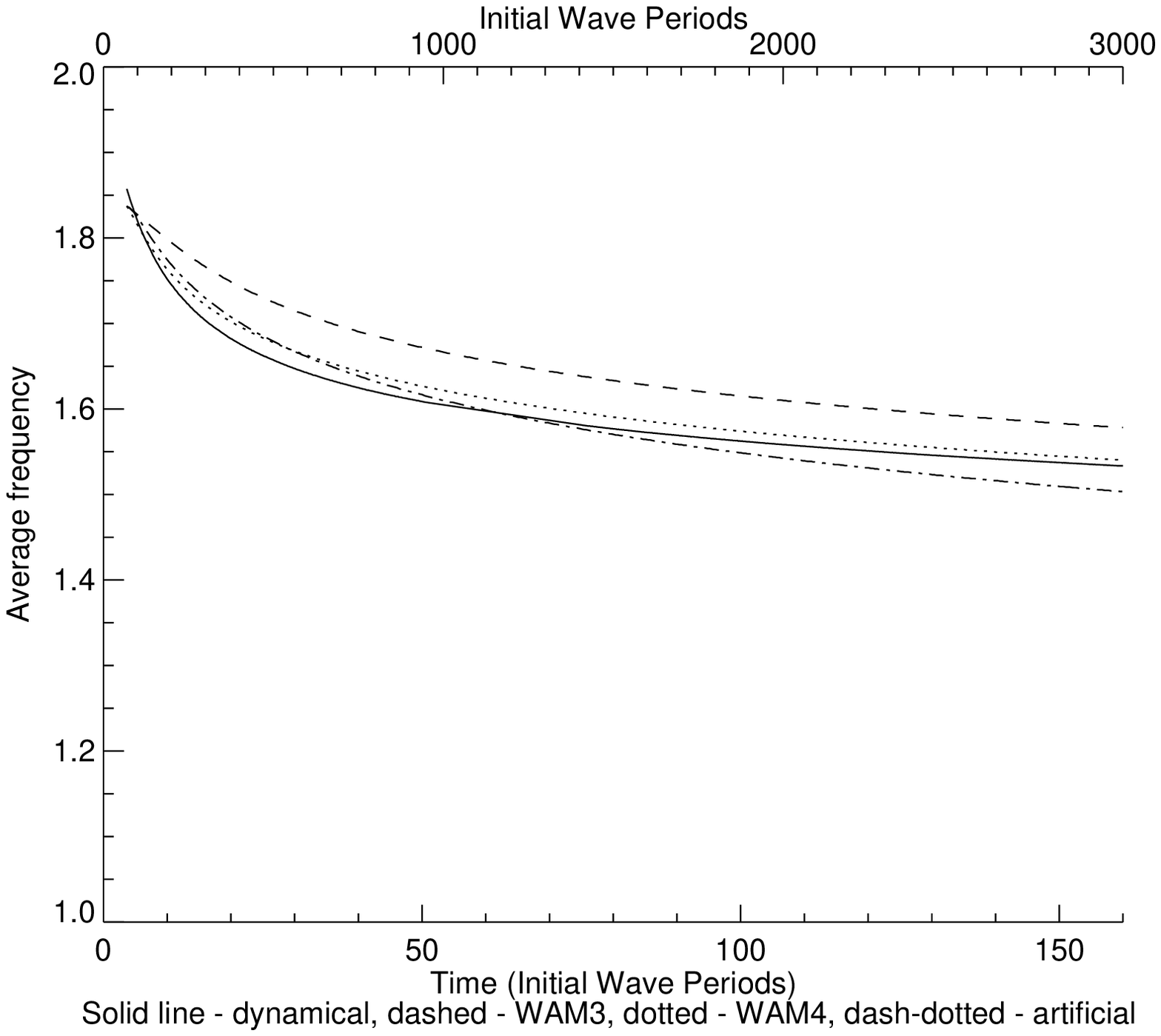} 
\caption{Mean wave frequency as a function of time. Solid line --- dynamical equations,
dashed-dotted line --- kinetic equation with artificial viscosity,
dashed line --- kinetic equation with {\it WAM3} damping term,
dotted line --- kinetic equation with {\it WAM4} damping term}\label{Freq} 
\end{figure}

Fig. \ref{Action}-\ref{Freq} demonstrate time evolution of main characteristics 
of the wave field: wave action, energy, characteristic slope and mean frequency. 


Fig.\ref{Slope_ArtVisc} should be specially commented. Here and further we define the characteristic slope as defined in Eq.~\ref{Steepness_definition}.
According to this definition of steepness for the classical Pierson-Moscowitz spectrum $\mu = 0.095$. Our initial 
steepness $\mu\simeq0.15$ exceeds this value essentially. 
 
\begin{figure}[htb] 
\centering 
\includegraphics[width=14.0cm]{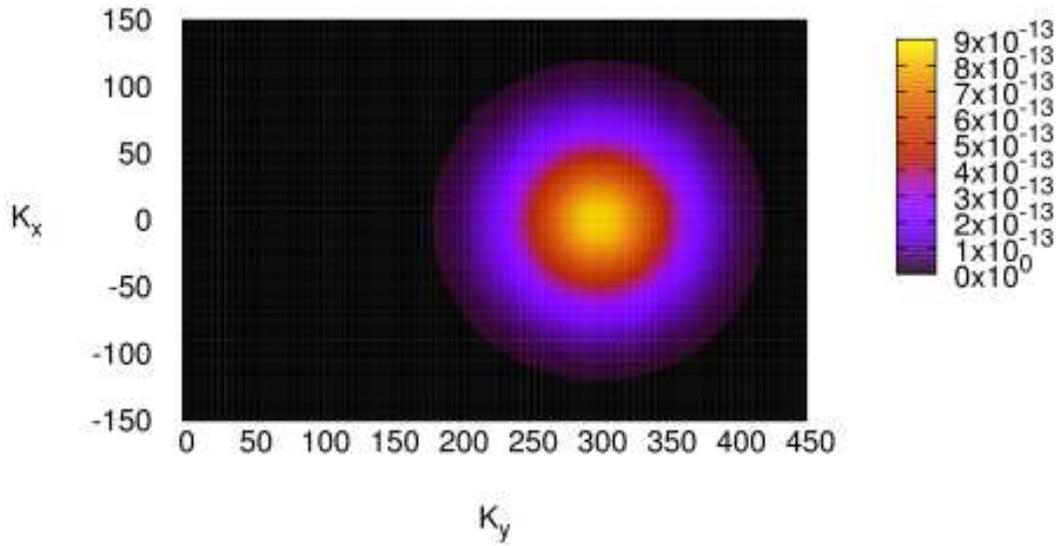}
\caption{\label{Spectrum_initial}Initial spectrum $|a_{\vec k}|^2$. $t=0$.} 
\end{figure}

\begin{figure}[htb] 
\centering 
\includegraphics[width=14.0cm]{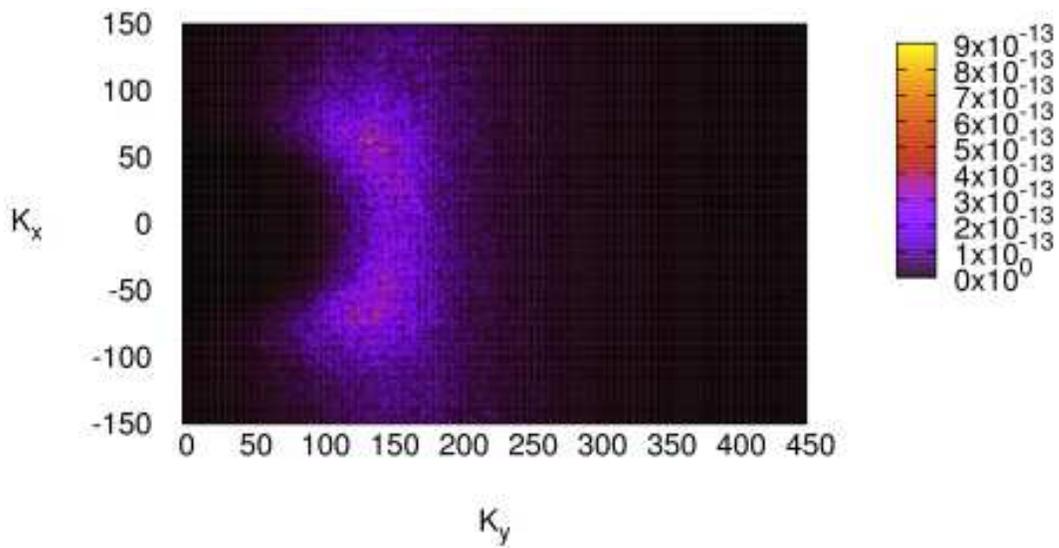}
\caption{\label{Spectrum_final}Final spectrum $|a_{\vec k}|^2$. 
$t=3378T_0$.} 
\end{figure}

Observed evolution of the spectrum can be conventionally separated in two phases. On the first stage we observe 
fast drop of wave action, slope and especially energy. Then the drop is stabilized, and we observe slow down-shift 
of mean frequency together with angular spreading. Level lines of smoothed 
spectra in the first and in the 
last moments of time are shown in 
Fig.\ref{Spectrum_initial}-\ref{Spectrum_final} 
 
\begin{figure}[htb] 
\centering 
\includegraphics[width=12.5cm,angle=0]{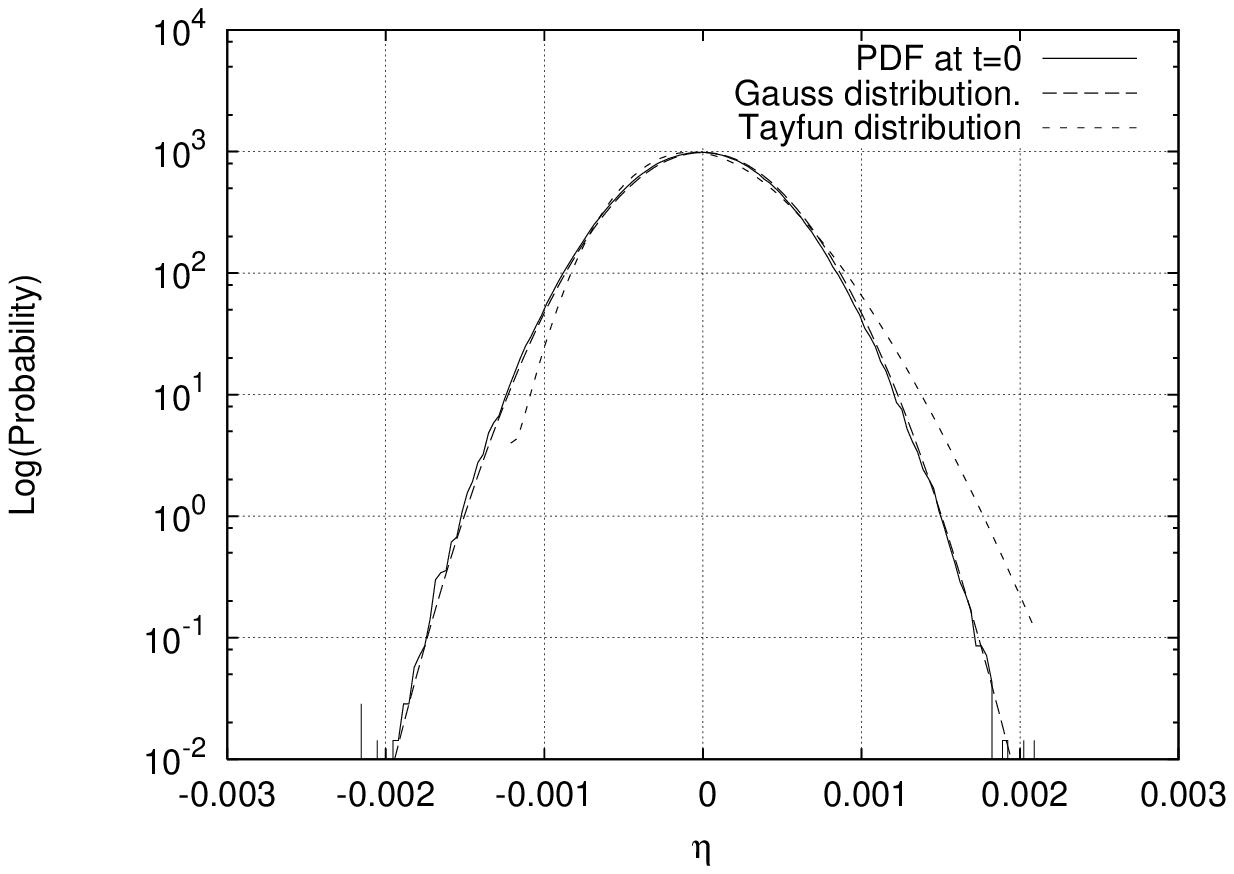} 
\caption{\label{PDF_eta_initial}PDF for the surface elevation $\eta$ at the 
initial moment of time. $t=0$.} 
\end{figure}

\begin{figure}[htb] 
\centering 
\includegraphics[width=12.5cm,angle=0]{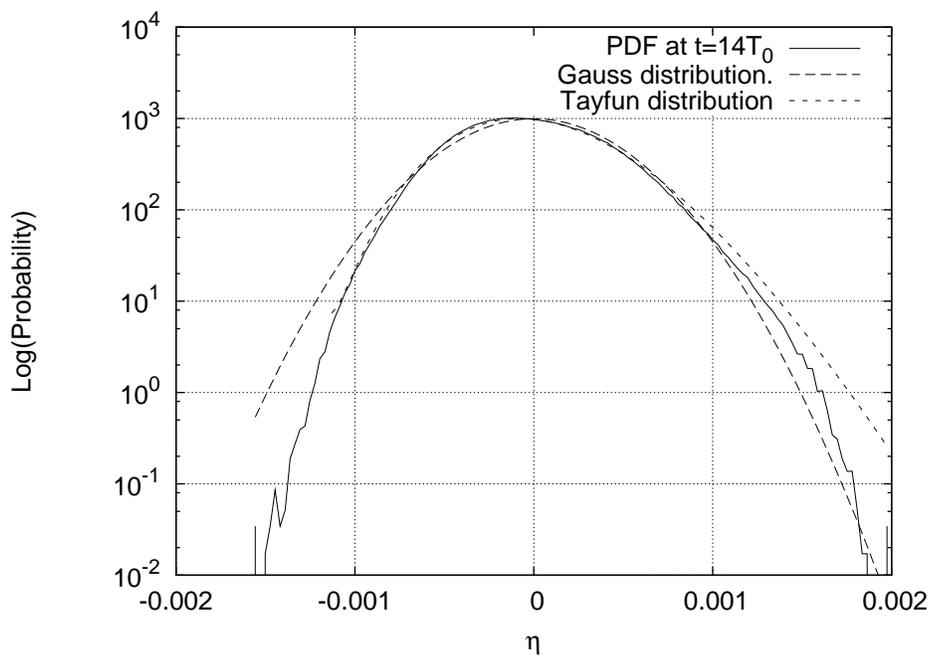} 
\caption{\label{PDF_eta_max_roughness}PDF for the surface elevation $\eta$ at the moment of
maximum surface roughness. $t\simeq14T_0$.} 
\end{figure}

\begin{figure}[htb] 
\centering 
\includegraphics[width=12.5cm,angle=0]{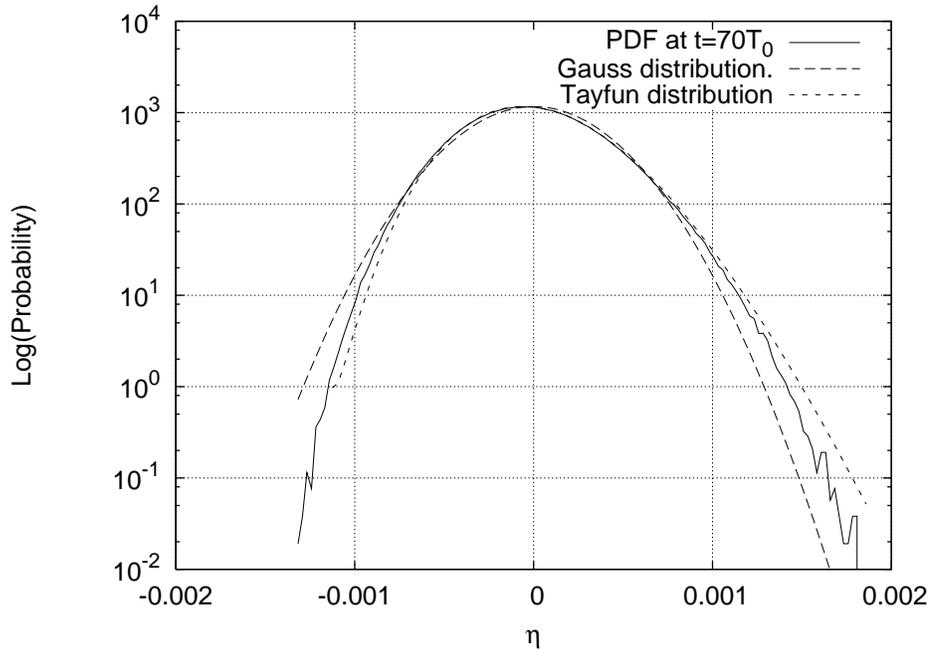} 
\caption{\label{PDF_eta_middle}PDF for the surface elevation $\eta$ at some 
middle moment of time. $t\simeq 67.1 T_0$.} 
\end{figure}

\begin{figure}[htb] 
\centering 
\includegraphics[width=12.5cm,angle=0]{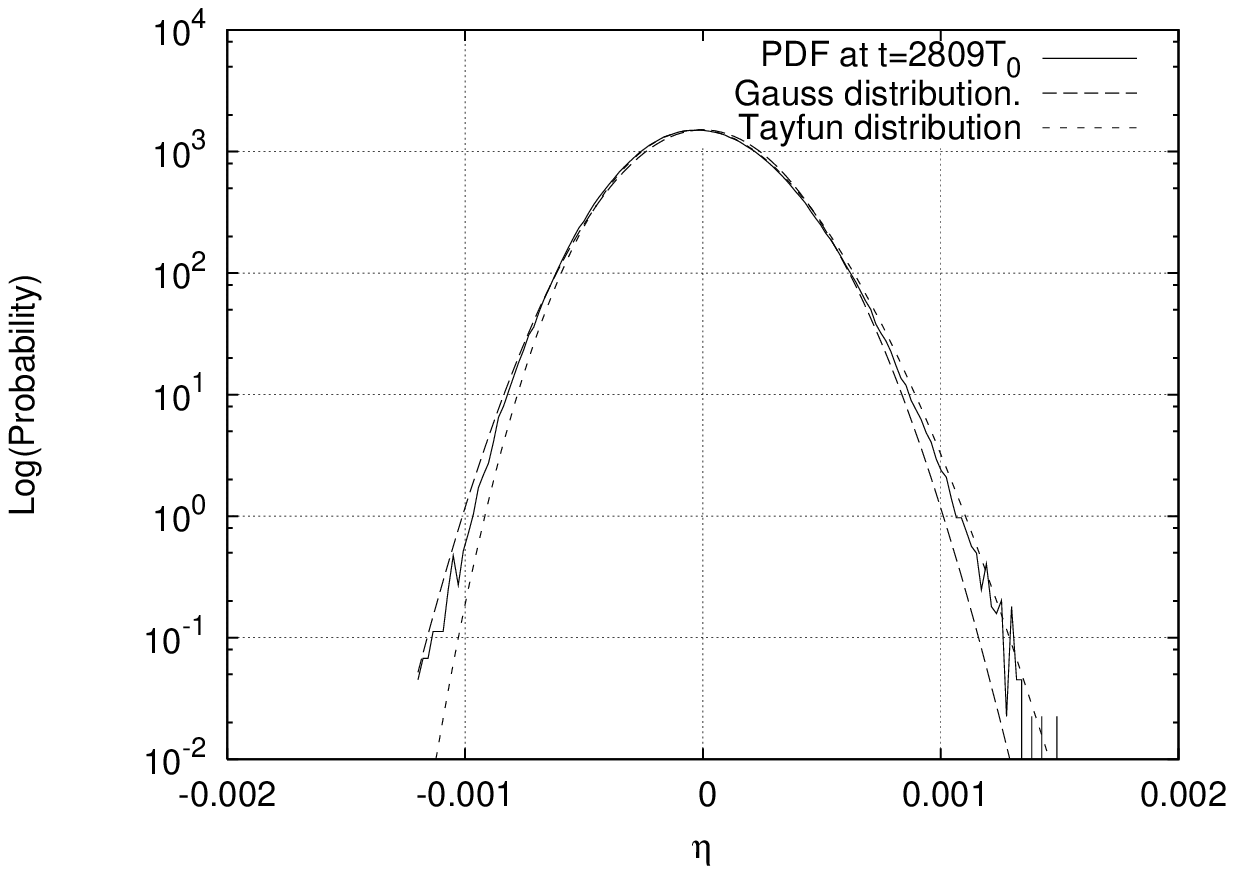} 
\caption{\label{PDF_eta_final}PDF for the surface elevation $\eta$ at the final 
moment of time. $t=3378T_0$.} 
\end{figure}

\begin{figure}[htb] 
\centering 
\includegraphics[width=12.5cm,angle=0]{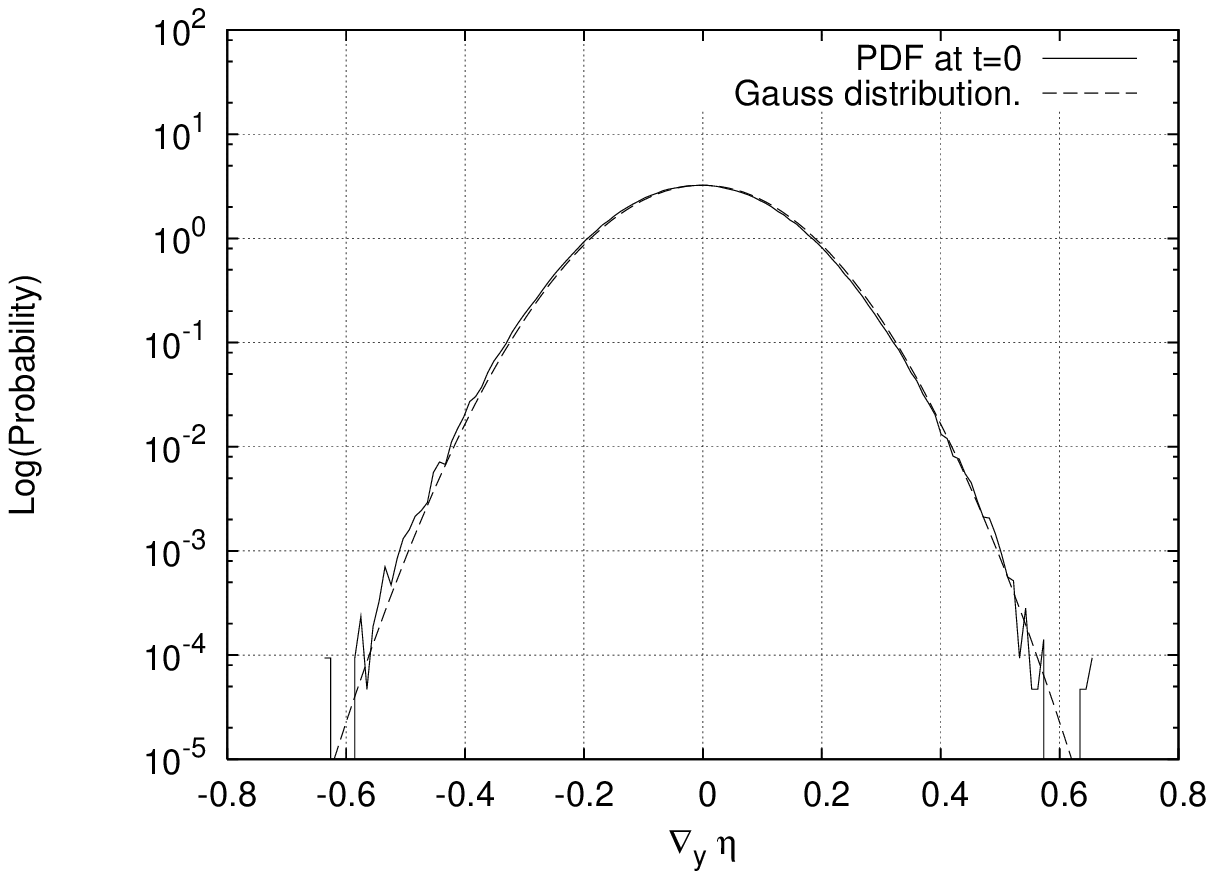} 
\caption{\label{GradY_initial}PDF for $(\nabla \eta)_y$ at the initial moment of 
time. $t=0$.} 
\end{figure}

\begin{figure}[htb] 
\centering 
\includegraphics[width=12.5cm]{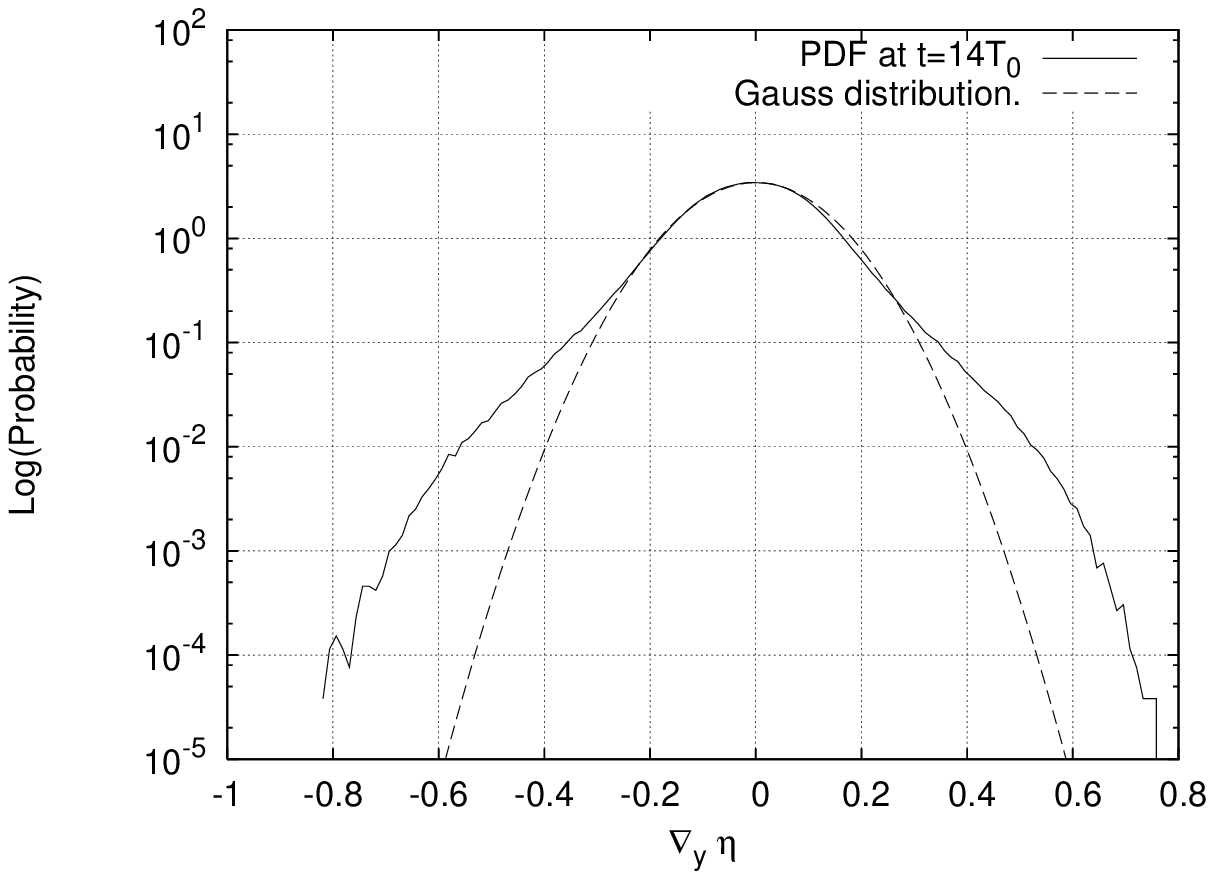} 
\caption{\label{GradY_max}PDF for $(\nabla \eta)_y$ at the moment of maximum 
surface roughness. $t\simeq14T_0$.} 
\end{figure}

\begin{figure}[htb] 
\centering 
\includegraphics[width=12.5cm,angle=0]{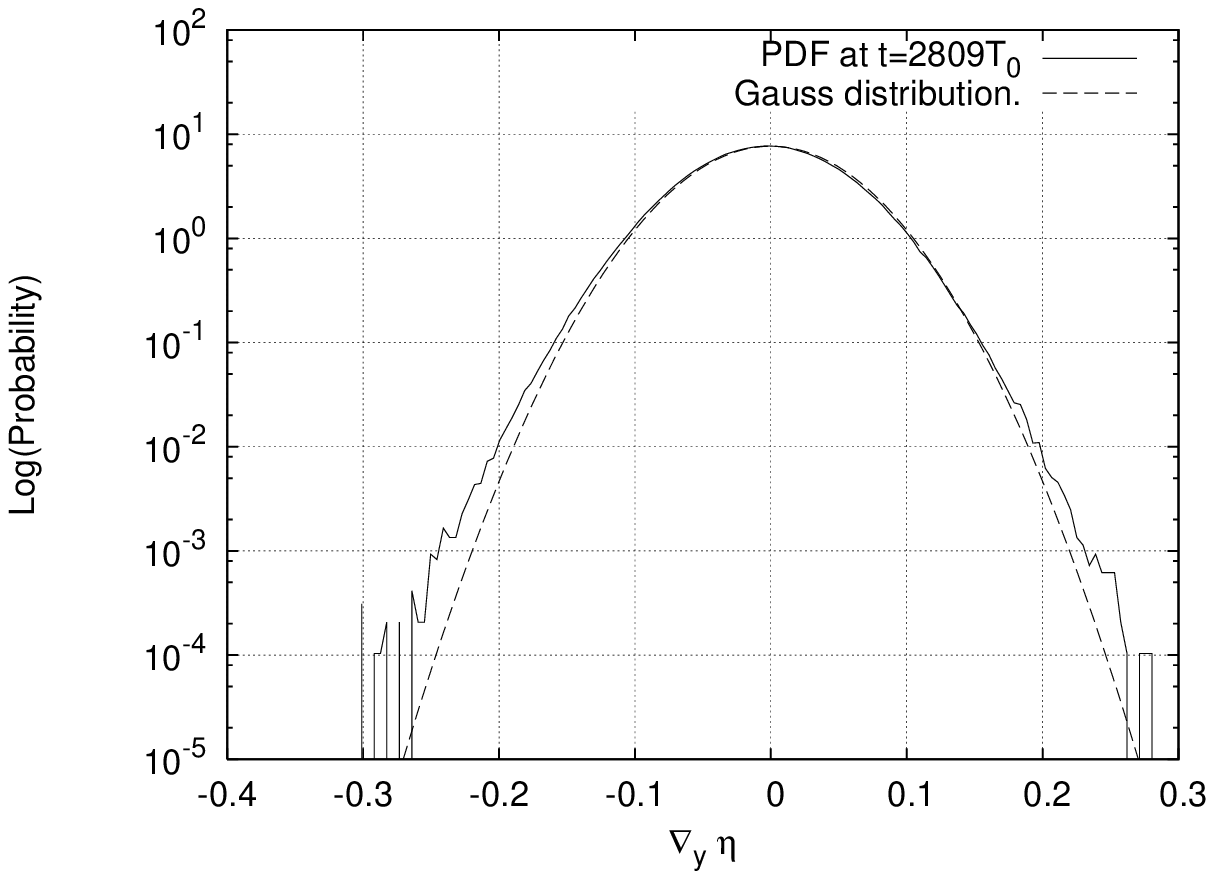} 
\caption{\label{GradY_final}PDF for $(\nabla \eta)_y$ at the final moment of 
time. $t=3378T_0$.} 
\end{figure}

Presence of two stages can be understood through the study of the Probability Distribution Functions (PDFs) of the elevations of the surface. In all figures we compare the distribution in experimental results with
Gaussian distribution and Tayfun distribution \cite{Tayfun1980}. The later case is just a first correction
to Gaussian distribution due to small nonlinearity. We used explicit form of Tayfun distribution following
\cite{Dysthe2005}.
In the initial moment of time the PDF is Gaussian (see Fig.\ref{PDF_eta_initial}). No nonlinear interaction is
involved, so Tayfun distribution does not fit at all. However, very soon intensive super-Gaussian tails appear
(see Fig.\ref{PDF_eta_max_roughness}). They are well described by Tayfun distribution.
Then tails decrease slowly (Fig.\ref{PDF_eta_middle}), and in the last moment of simulation, when characteristics of 
the sea are close to Peirson-Moscowitz, the statistics is close to Gaussian again (Fig.\ref{PDF_eta_final}). Moderate tails do exist and, in a good agreement with Tayfun correction, troughs are more probable than crests which
in turn can be much larger in absolute value. 
PDF for $\eta_y$ --- longitudinal gradients in the first moments of time 
is Gaussian (Fig.\ref{GradY_initial}). Then in a very short period of time 
strong non-Gaussian tails appear and reach their 
maximum at $t\simeq14T_0$ (Fig.\ref{GradY_max}). Here $T_0 = 2\pi/\sqrt{k_0}$ 
--- period of initial leading wave. Since this 
moment the non-Gaussian tails decrease. In the last moment of simulation they 
are essentially reduced(Fig.\ref{GradY_final}).

Fast growth of non-Gaussian tails can be explained by formation of coherent
harmonics. Indeed, $14T_0 \simeq 2\pi/(\omega_0 \mu)$ is a characteristic time of
nonlinear processes due to quadratic nonlinearity. Note that the fourth harmonic
in our system is strongly decaying, hence we cannot see real white caps.
 
\begin{figure}[htb] 
\centering 
\includegraphics[width=12.5cm]{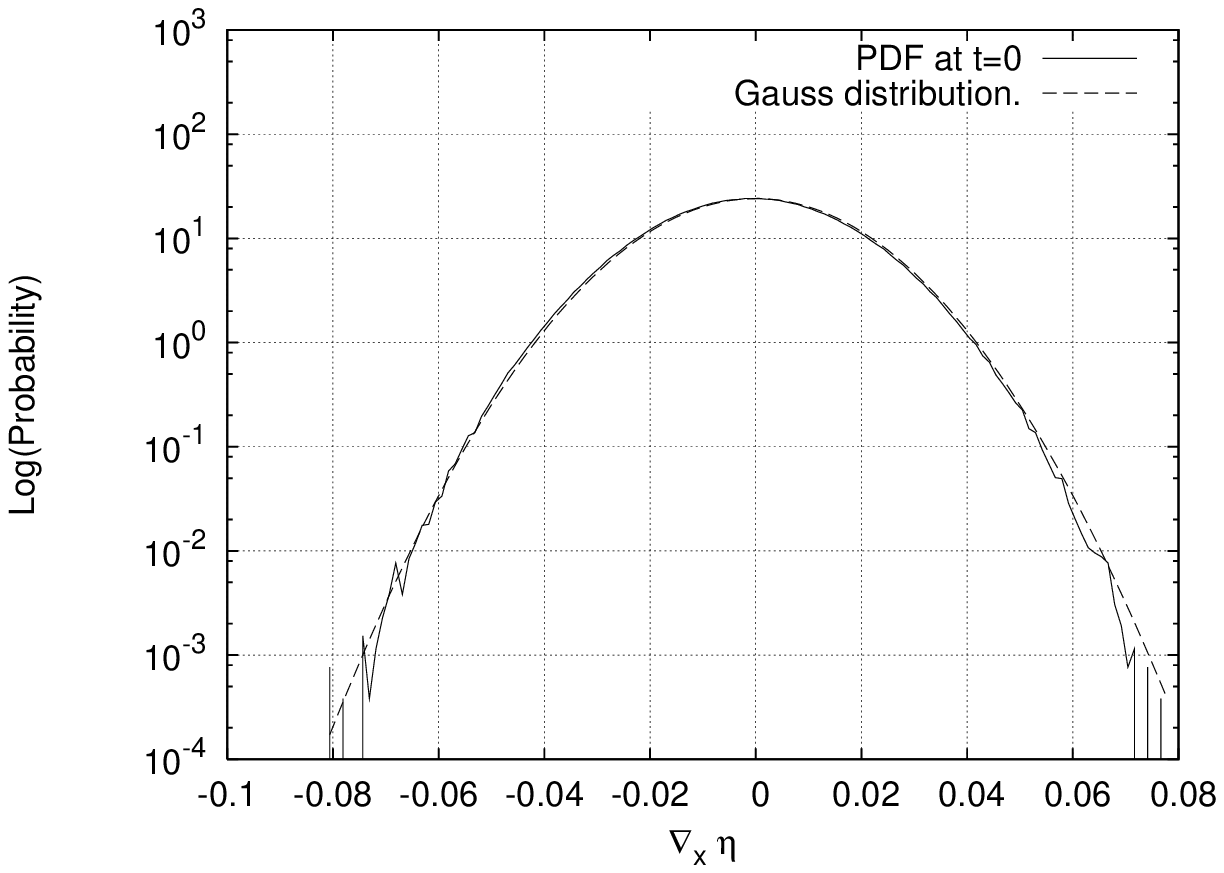} 
\caption{\label{GradX_initial}PDF for $(\nabla \eta)_x$ at the initial moment of 
time. $t=0$.} 
\end{figure}

\begin{figure}[htb] 
\centering 
\includegraphics[width=12.5cm]{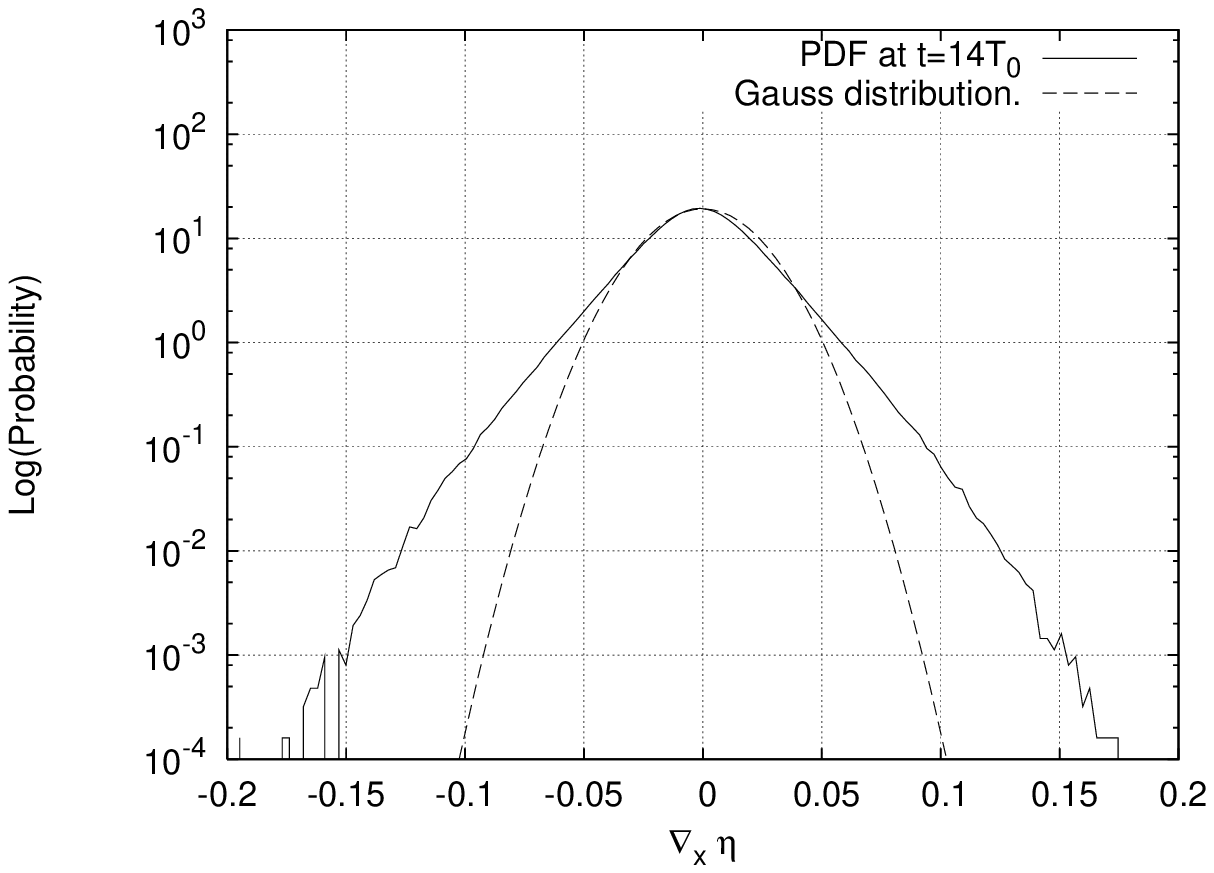} 
\caption{\label{GradX_max}PDF for $(\nabla \eta)_x$ at the moment of maximum 
surface roughness. $t\simeq14T_0$.} 
\end{figure}

\begin{figure}[htb] 
\centering 
\includegraphics[width=12.5cm]{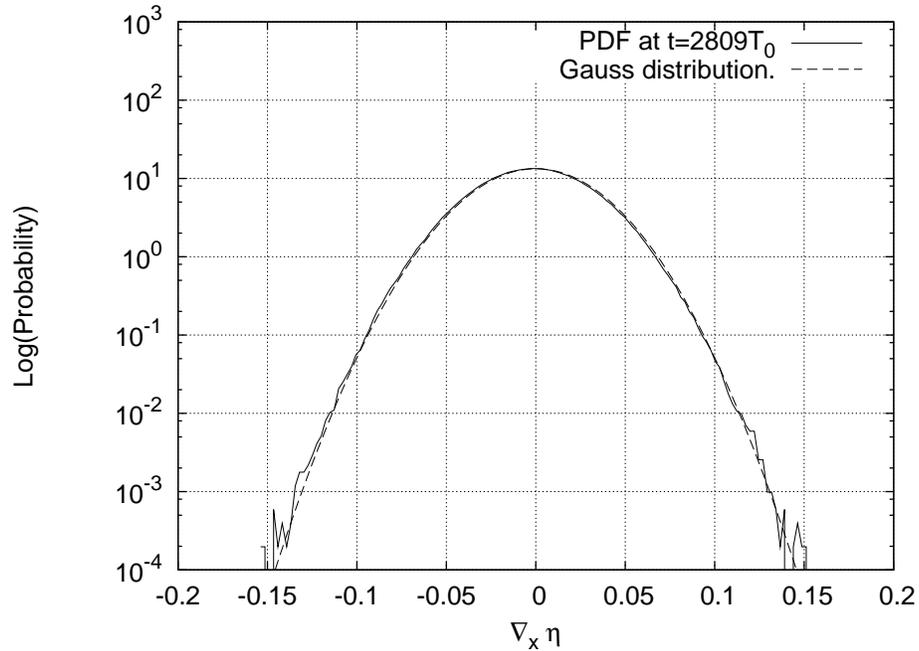} 
\caption{\label{GradX_final}PDF for $(\nabla \eta)_x$ at the final moment of 
time. $t=3378T_0$.} 
\end{figure}

\begin{figure}[htb] 
\centering 
\includegraphics[width=14.0cm]{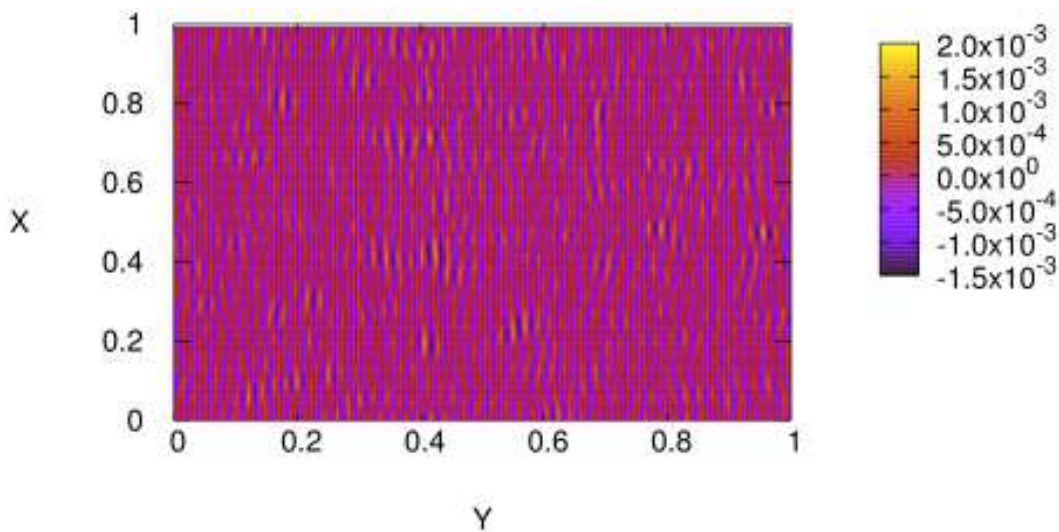}
\caption{\label{Eta_surf_initial} Surface elevation at the initial moment of 
time. $t=0$.} 
\end{figure}

\begin{figure}[htb] 
\centering 
\includegraphics[width=14.0cm]{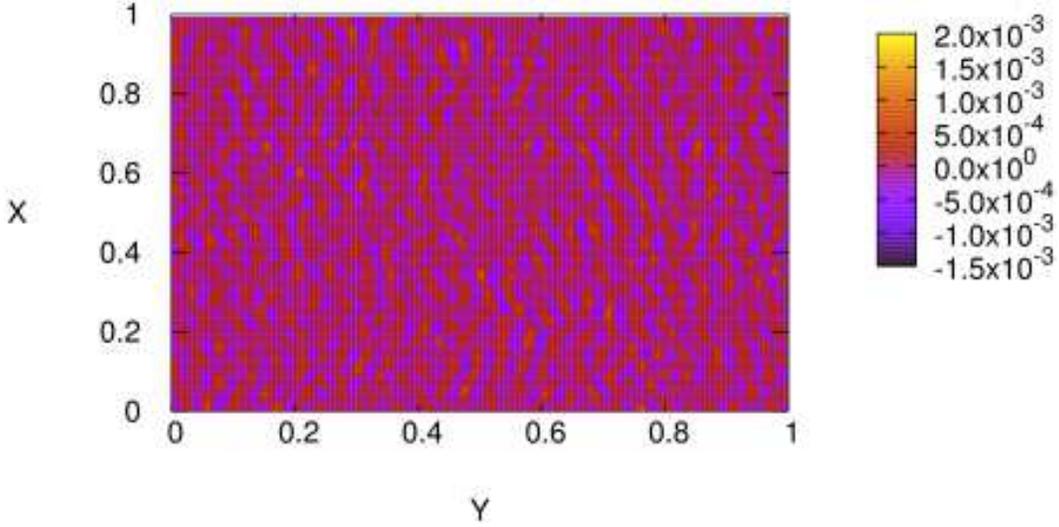}
\caption{\label{Eta_surf_final} Surface elevation at the final moment of time. 
$t=3378T_0$.} 
\end{figure}

\begin{figure}[htb] 
\centering 
\includegraphics[width=14.5cm]{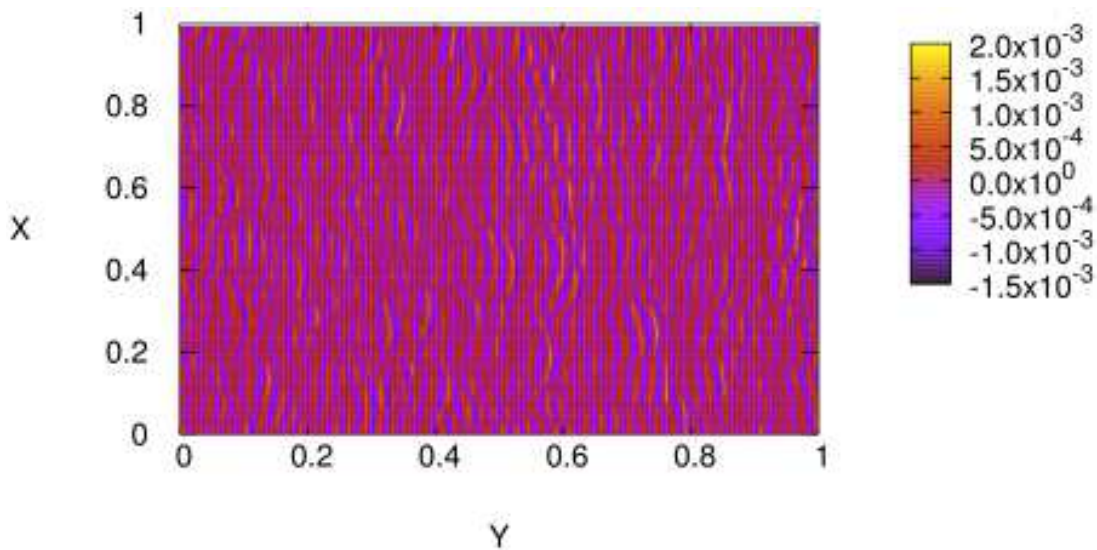}
\caption{\label{Eta_surf_max} Surface elevation at the moment of maximum 
surface roughness. $t\simeq14T_0$. Gradients are more conspicuous.} 
\end{figure}

Figures \ref{GradX_initial}-\ref{GradX_final} present PDFs for gradients in the 
orthogonal direction. 
\clearpage 
Figures \ref{Eta_surf_initial},\ref{Eta_surf_final} present snapshots of the 
surface in the initial and final moments 
of simulation. Fig.\ref{Eta_surf_max} is a snapshot of the surface in the moment 
of maximal roughness $T=4.94\simeq14T_0$. 
 
\section{Statistical numerical experiment} 
 
\subsection{Numerical model for Hasselmann Equation} 
 
Numerical integration of kinetic equation for gravity waves on deep water 
(Hasselmann equation) was the subject of considerable efforts for last three 
decades. 
The ``ultimate goal'' of the effort -- creation of the operational wave model 
for 
wave forecast based on direct solution of the Hasselmann equation -- happened to 
be an extremely difficult computational problem due to mathematical complexity 
of 
the $S_{nl}$ term, which requires calculation of a three-dimensional integral at 
every moment of time. 
 
Historically, numerical methods of integration of kinetic equation for gravity 
waves exist in two ``flavors''. 
 
The first one is associated with works \cite{Hasselmann1985}, 
\cite{Dungey1985}, 
\cite{Masuda1981}, \cite{Masuda1986}, \cite{Lavrenov1998} and 
\cite{Polnikov2001}, 
and is based on transformation of 6-fold 
into 3-fold integrals using $\delta$-functions. Such transformation leads to 
appearance of integrable singularities, which creates additional difficulties 
in calculations of the $S_{nl}$ term. 
 
The second type of models has been developed in works of \cite{Webb1978} and 
\cite{Resio1982}, \cite{Resio1991} and is currently known as Resio-Tracy model. 
It uses direct calculation of resonant quadruplet contribution 
into $S_{nl}$ integral, based on the following property: given two fixed vectors 
$\vec{k}, \vec{k_1}$, another two $\vec{k_2}, \vec{k_3}$ are uniquely defined by 
the point ``moving'' along the resonant curve -- locus. 
 
Numerical simulation in the current work was performed with the help of 
modified version of the second type algorithm. Calculations were made on the 
grid $71 \times 36$ points in the frequency-angle domain $[\omega, \theta]$ with 
exponential distribution of points in the frequency domain and uniform 
distribution of points in the angle direction. 
 
To date, Resio-Tracy model suffered rigorous testing and is currently used with 
high degree of trustworthiness: it was tested with respect to  motion integrals 
conservation in the ``clean'' tests, wave action conservation in wave spectrum 
down-shift, realization of self -- similar solution in ``pure swell'' and ``wind 
forced'' 
regimes (see \cite{Pushkarev2000HE}, \cite{Pushkarev2003}, \cite{Badulin2005}). 
 
Description of scaling procedure from dynamical equations to Hasselman kinetic 
equation 
variables is presented in Appendix \ref{APP:FromDynamics2Hasselmann}. 
 
\subsection{Statistical model setup} 
 
The numerical model used for solution of the Hasselmann equation has been 
supplied with the damping term in three different forms: 
\begin{enumerate} 
\item{} Pseudo-viscous high frequency damping (\ref{Pseudo_Viscous_Damping}) 
used in dynamical equations; 
\item{} {\it WAM3} viscous term; 
\item{} {\it WAM4} viscous term; 
\end{enumerate} 
Two last viscous terms are the ``white-capping'' terms, describing energy 
dissipation by surface waves due to white-capping, as used in $WAM$
wave forecasting models (in $SWAN$ model the used term has different but close tunable parameters),
see \cite{SWAN}: 
\begin{eqnarray} 
\label{WAMdissipation} 
\gamma_{\vec{k}} = C_{ds}  \tilde{\omega} \frac{k}{\tilde{k}} \left((1-\delta)+
\delta\frac{k}{\tilde{k}}\right)\left(\frac{\tilde{S}}{\tilde{S}_{pm}}\right)^p 
\end{eqnarray} 
where $k$ and $\omega$ are wave number and frequency, tilde denotes mean value; 
$C_{ds}$, $\delta$ and $p$ 
are tunable coefficients; $S=\tilde{k}\sqrt{H}$ is the overall steepness; 
$\tilde{S}_{PM}=(3.02\times 10^{-3})^{1/2}$ 
is the value of $\tilde{S}$ for the Pierson-Moscowitz spectrum (note that the 
characteristic steepness 
$\mu = \sqrt{2} S$). It is worth to note that according to \cite{Yang}
theoretical value of steepness for Pierson-Moscovitz spectrum is
$S_{PM} \simeq (4.57\times 10^{-3})^{1.2}$. It gives us $\mu\simeq 0.095$.
 
Values of tunable coefficients for {\it WAM3} case are: 
\begin{equation} 
C_{ds} = 2.36 \times 10^{-5},\,\,\,\delta=0,\,\,\,p=4 
\end{equation} 
and for {\it WAM4} case are: 
\begin{equation} 
C_{ds} = 4.10 \times 10^{-5},\,\,\,\delta=0.5,\,\,\,p=4 
\end{equation} 
 
In all three cases we used as initial data smoothed (filtered) spectra 
(see Fig.\ref{SmoothedDistribution}) 
obtained in the dynamical run at the time $T_{*} = 3.65 min = 24.3 \simeq 67.1 
T_0$, which can be considered 
as a moment of the end of the first ``fast'' stage of spectral evolution. 
 
The natural question stemming in this point, is why calculation of the dynamical 
and Hasselmann model cannot 
be started from the initial conditions (\ref{Dynamic_initial_conditions}) 
simultaneously? 
 
There are good reasons for that: 
 
As it was mentioned before, the time evolution of the initial conditions 
(\ref{Dynamic_initial_conditions}) in presence of the damping term  can be 
separated in two stages: relatively fast total energy drop in the beginning of 
the evolution and succeeding relatively slow total energy decrease as a function 
of time, see Fig.\ref{Energy}. We explain this phenomenon by existence of the 
effective channel of the energy dissipation due to strong nonlinear effects, 
which can be associated with the white-capping as we mentioned in the introduction. 
 
We have started with relatively steep waves $\mu\simeq 0.167$. As we see, at 
that steepness white-capping is the leading effect. This fact is confirmed by 
numerous field and laboratory experiments. From the mathematical view-point, the 
white-capping is formation of coherent structures -- strongly correlated 
multiple harmonics. The spectral peak is posed in our experiments initially at 
$k\simeq 300$, while the edge of the damping area $k_d \simeq 1024$. Hence, only 
the second and the third harmonic can be developed, while higher harmonics are 
suppressed by strong dissipation. Anyway, even formation of the second and 
the third harmonic is enough to create intensive non-Gaussian tail of the $PDF$ 
for longitudinal gradients. This process is very fast. In the moment of time 
$T=14 T_0$ we see fully developed tails. Relatively sharp gradients mimic 
formation of white caps. Certainly, the ``pure'' Hasselmann equation is not 
applicable on this early stage of spectral evolution, when energy intensively 
dissipates. 
 
As steepness decreases and spectral maximum of the swell down-shifts, the 
efficiency of such mechanism of energy absorption becomes less important. 
When the steepness value drops down to $\mu \simeq 0.1$, at approximately 
$T\simeq 280 T_0$, the white-capping is negligibly small. Therefore, we decided 
to start comparison between deterministic and statistical modeling in some 
intermediate moment of time $T\simeq 67.1 T_0$. 
 
\section{Comparison of deterministic and statistical experiments.} 
\subsection{Statistical experiment with pseudo-viscous damping term.} 
The first series of statistical experiments has been performed  with pseudo-viscous damping term (\ref{Pseudo_Viscous_Damping}). 
 
Fig.\ref{Action} -- \ref{Freq} show total wave action, total 
energy, mean wave slope and mean wave frequency as the functions of time. 
 
\begin{figure}[ht] 
\centering
\includegraphics[width=8.5cm,angle=90]{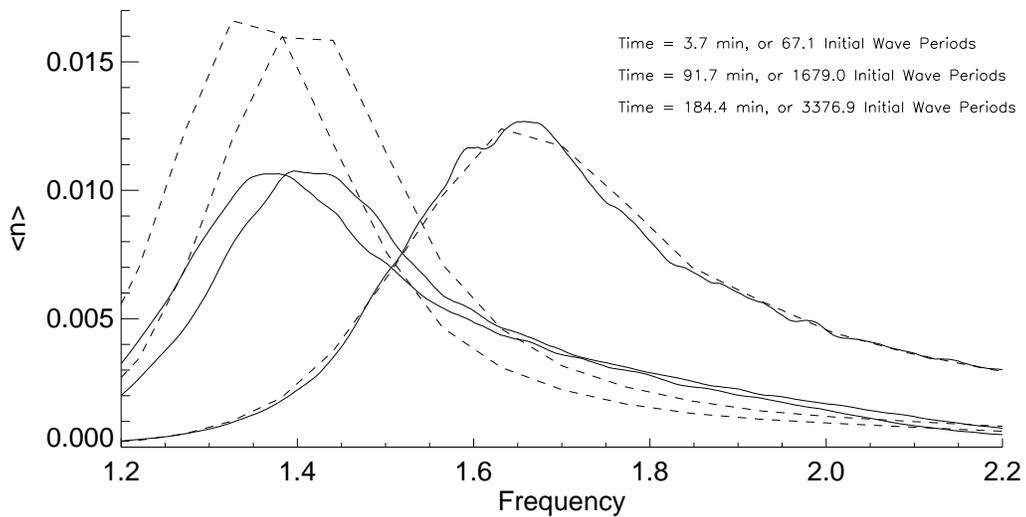} 
\caption{Angle-averaged spectrum as a function of time for dynamical and Hasselmann equations for artificial viscosity case.}\label{AngleAver_ArtVisc} 
\end{figure}

Fig.\ref{AngleAver_ArtVisc} shows the time evolution of angle-averaged wave 
action spectra as the functions of frequency for dynamical and Hasselmann 
equations. We see similar down-shift of the spectral maximum both in dynamic and 
Hasselmann equations. The correspondence of the spectral maxima amplitudes is not good at all. 
 
It is quite obvious that the influence of the artificial viscosity is not strong 
enough to reach  the correspondence of two models. 
 
\subsection{Statistical experiments with {\it WAM3} damping term} 
The second series of statistical experiments has been done for the choice of {\it WAM3}
damping term.

The temporal behavior of total wave action, energy and average wave slope 
(see Fig.\ref{Action} -- \ref{Freq}) for {\it WAM3} damping term is in better 
correspondence with dynamical model, than in the case of artificial viscosity term. 
For initial $50 \, min$ duration of the experiment  we observe decent correspondence 
between dynamical and Hasselmann equations. For longer time the {\it WAM3} model, however, 
deviates from the dynamical model significantly.

\begin{figure}[ht] 
\centering
\includegraphics[width=8.5cm,angle=90]{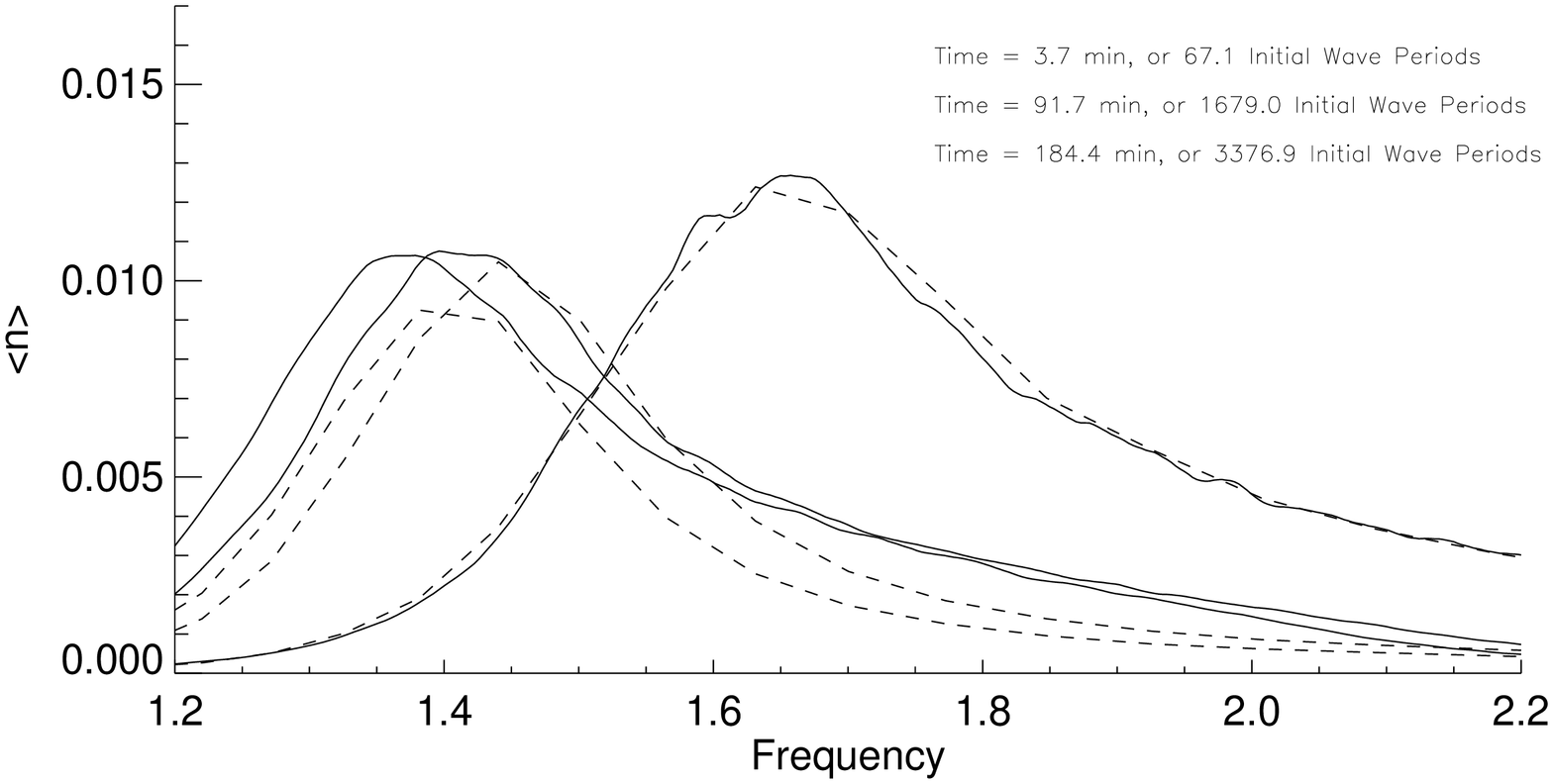} 
\caption{Angle-averaged spectrum as a function of time for dynamical and Hasselmann equations a function of time for {\it WAM3} case.}\label{AngleAver_WAM2} 
\end{figure}

As in the artificial viscosity case, the angle-averaged wave action spectra as the 
function of frequency exhibit distinct down-shift of the spectral maxima for 
both dynamical and Hasselmann equations (see Fig.\ref{AngleAver_WAM2}). 
Correspondence between time evolution of the amplitudes of the spectral maxima is 
also much better for {\it WAM3} choice of damping, than for artificial viscosity case. 
 
Presumably, {\it WAM3} damping term underestimate the effects of real damping at the very
beginning of the evolution (when the effects of white capping are relatively important), and
overestimates them on later stages of swell evolution.

\subsection{Statistical experiments with {\it WAM4} damping term} 
 The final third series of experiments have been done for the choice of {\it WAM4} damping term.

Fig.\ref{Action}--\ref{Freq} show temporal evolution of the total wave action, 
total energy, mean wave slope and mean wave frequency, which are divergent in time in 
this case. 
 
\begin{figure}[ht]
\centering
\includegraphics[width=8.5cm,angle=90]{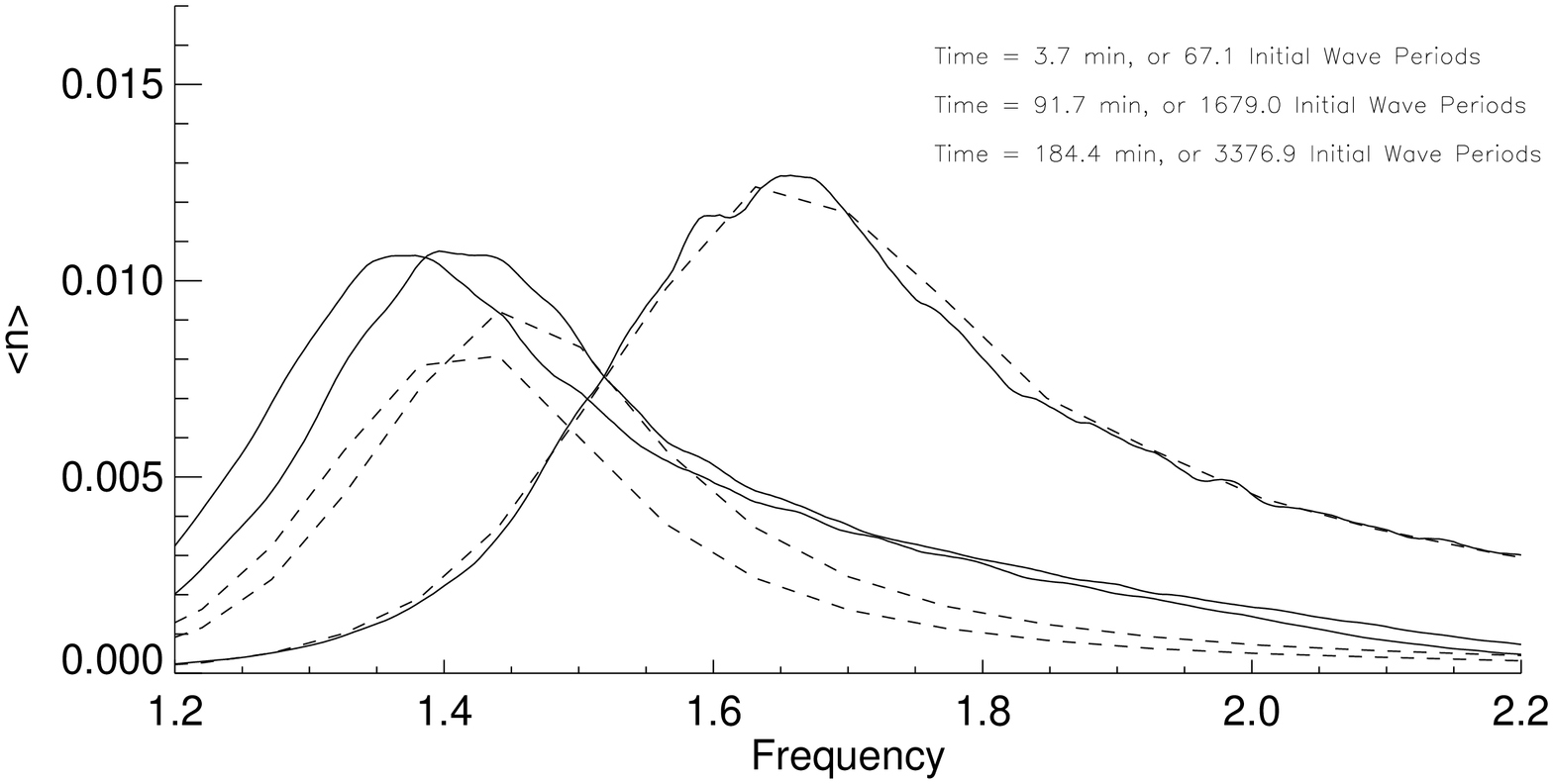} 
\caption{Angle-averaged spectrum as a function of time for dynamical and Hasselmann equations a function of time for {\it WAM4} case.}\label{AngleAver_WAM1} 
\end{figure}

Fig.\ref{AngleAver_WAM1} show time evolution of angle-averaged wave action 
spectra as the functions of frequency for dynamical and Hasselmann equations. 
While as in the artificial viscosity and {\it WAM3} cases we also observe distinct 
down-shift of the spectral maxima, the correspondence of the time evolution of 
the amplitudes of the spectral maxima is worse than in {\it WAM3} case. 
 
This observation is especially surprising in view of the fact that historically {\it WAM4} 
damping has been invented as an improvement to {\it WAM3} damping term. It is quite obvious that
{\it WAM4} damping is too strong for description of the reality at all stages of the swell evolution. 
 
\section{Down-shift and angular spreading} 
The major process of time-evolution of the swell is frequency down-shift. During 
$T=933 T_0$ the mean frequency has been decreased from $\omega_0 = 2$ to 
$\omega_1 = 0.6$. On the last stage of the process, the mean frequency slowly 
decays as 
\begin{equation} 
\label{Law1} 
<\omega> \sim t^{-0.067} \simeq t^{-1/15} 
\end{equation} 
The Hasselmann equation has self-similar solution, describing the evolution of 
the swell $n(\vec{k},t) \sim t^{4/11} F\left(\frac{\vec{k}}{t^{2/11}}\right)$ 
(see \cite{Pushkarev2003}, \cite{Badulin2005}). For this solution 
\begin{equation} 
\label{Law2} 
<\omega> \sim t^{-1/11} 
\end{equation} 
The difference between (\ref{Law1}) and (\ref{Law2}) can be explained as 
follows. What we observed, is not a self-similar behavior. Indeed, a 
self-similarity presumes that the angular structure of the solution is constant 
in time. Meanwhile, we observed intensive angular spreading of the initially 
narrow in angle, almost one-dimensional wave spectrum.

\begin{figure}[htb]
\centering
\includegraphics[width=7.5cm,angle=0]{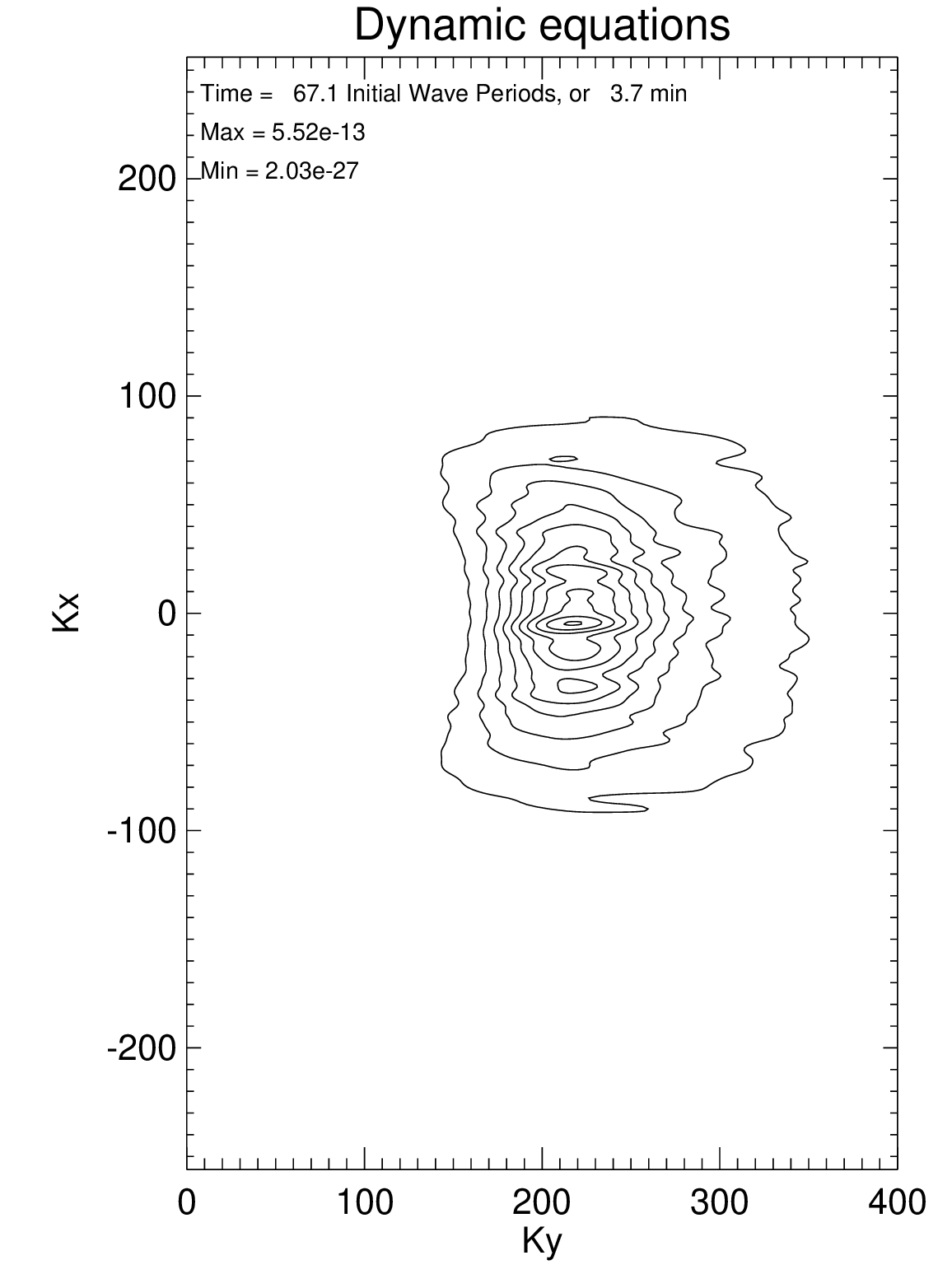} 
\includegraphics[width=7.5cm,angle=0]{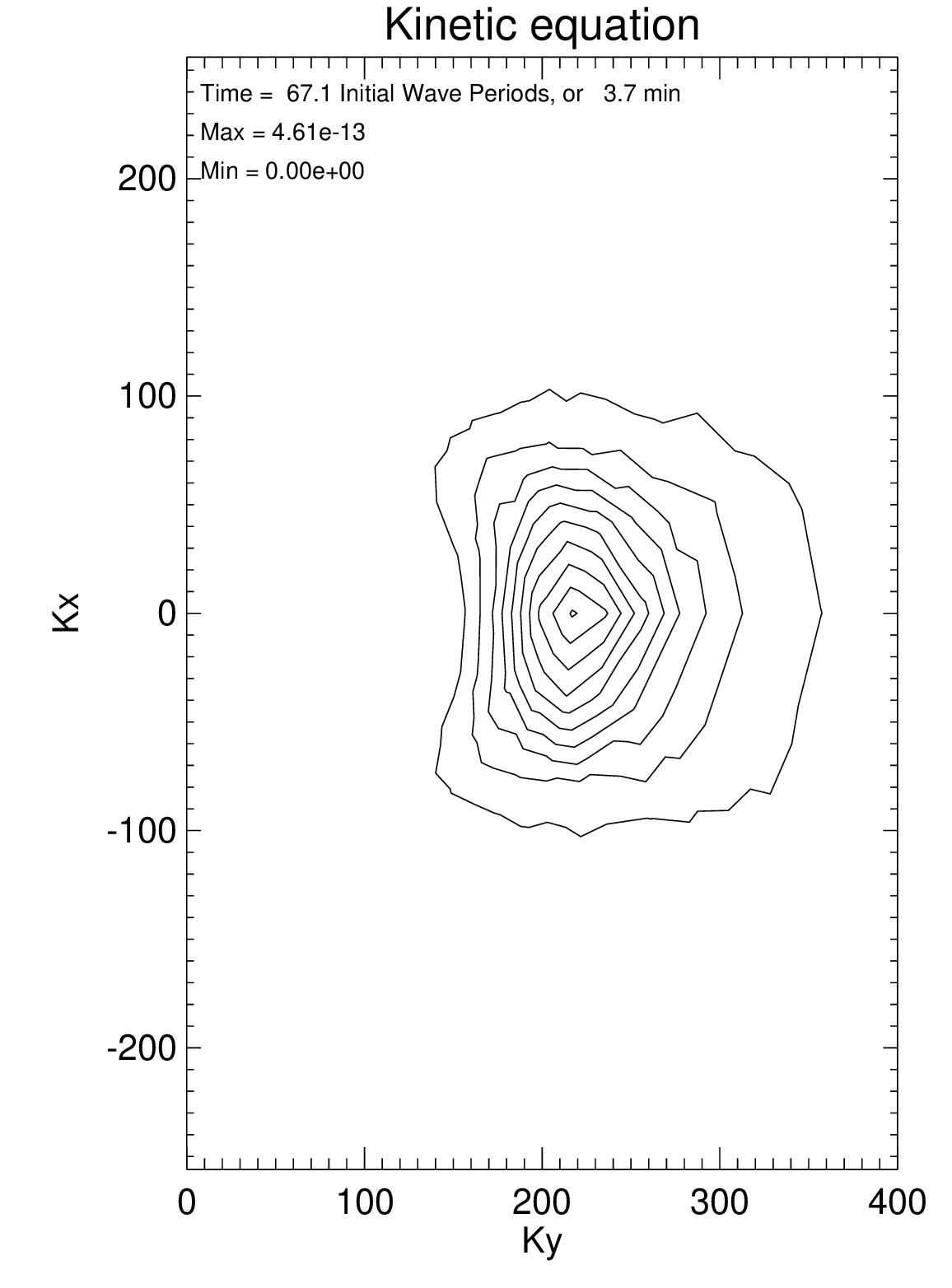} 
\caption{\label{DynStart}\label{KinStart}Level lines of the dynamic (left) and the kinetic (right) spectra at $t=67.1\,T_0$.}
\end{figure}

\begin{figure}[htb]
\centering
\includegraphics[width=7.5cm,angle=0]{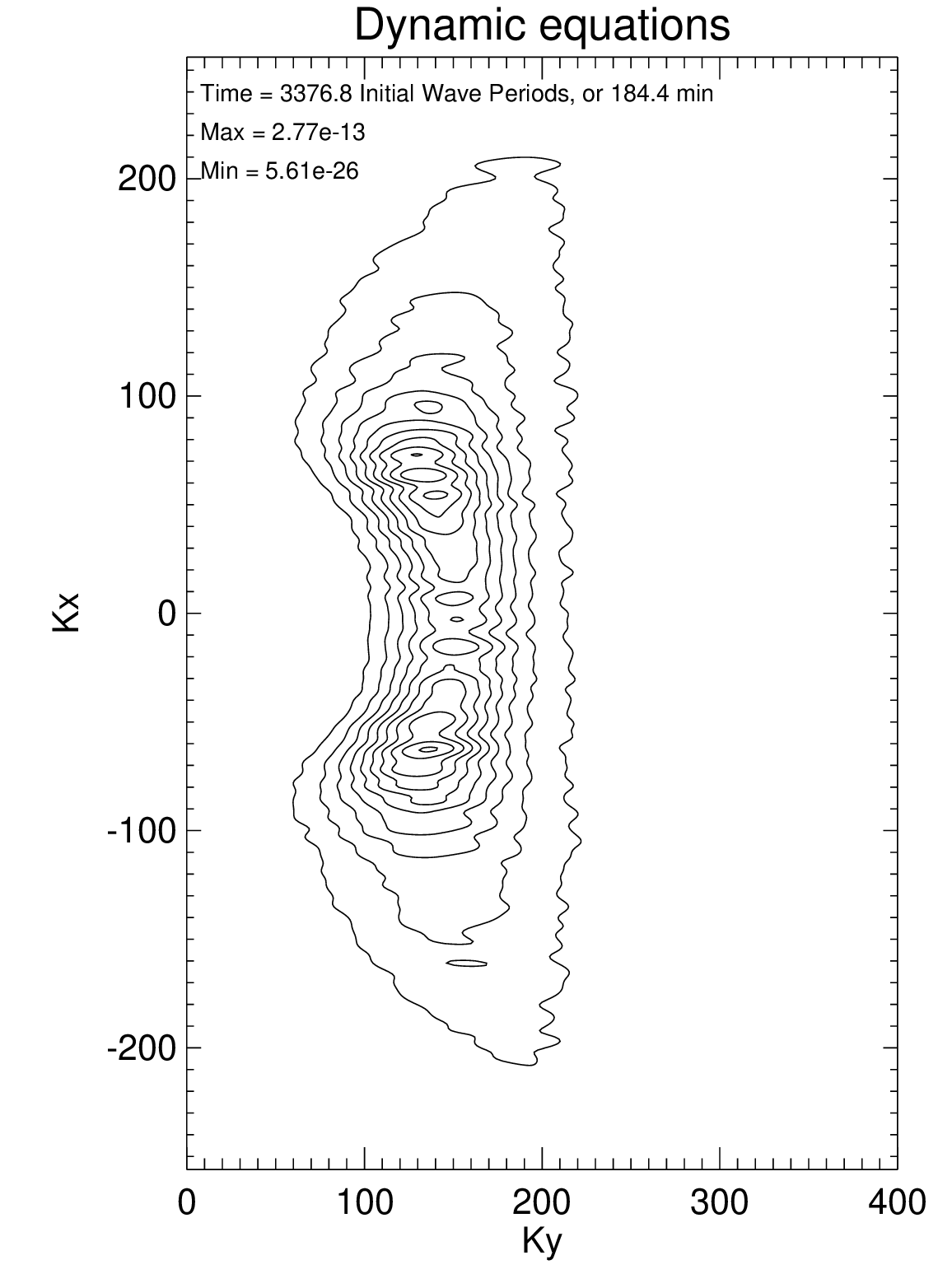}
\includegraphics[width=7.5cm,angle=0]{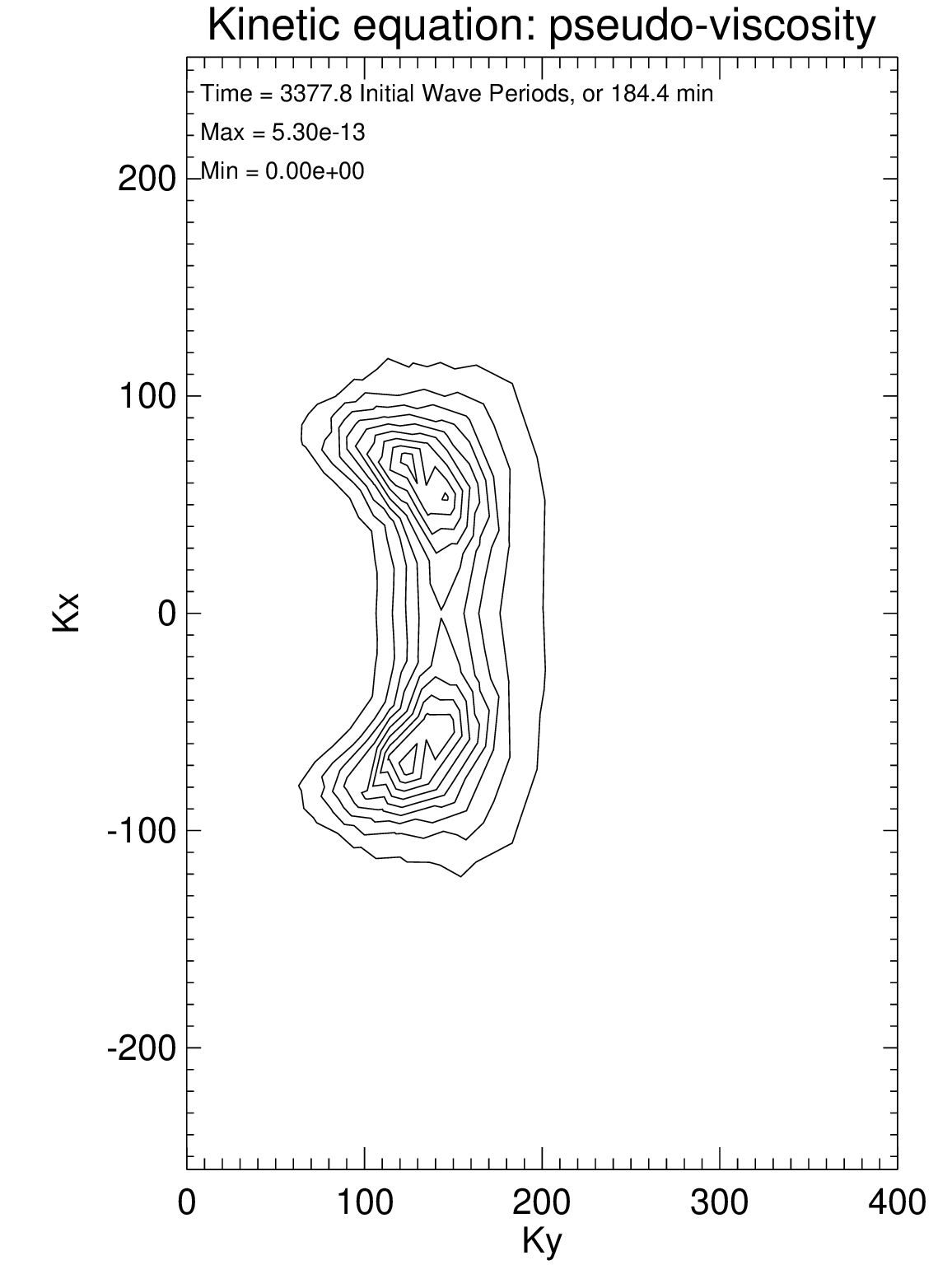}
\caption{\label{DynEnd}\label{KinEnd}Level lines of the dynamic (left) and the kinetic (right) spectra at $t=3378.2\,T_0$.}
\end{figure}

\begin{figure}[htb]
\centering
\includegraphics[width=7.5cm,angle=0]{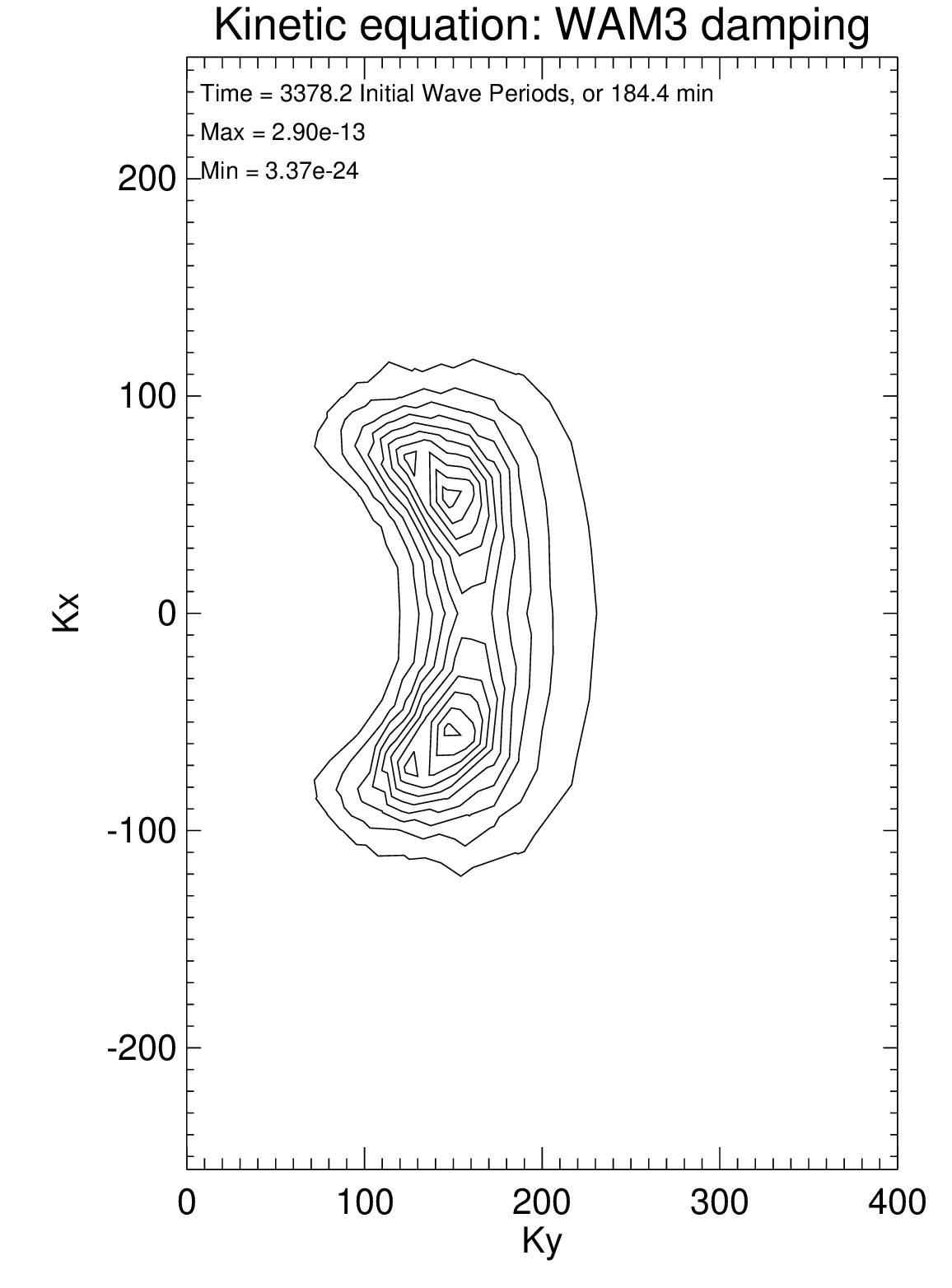}
\includegraphics[width=7.5cm,angle=0]{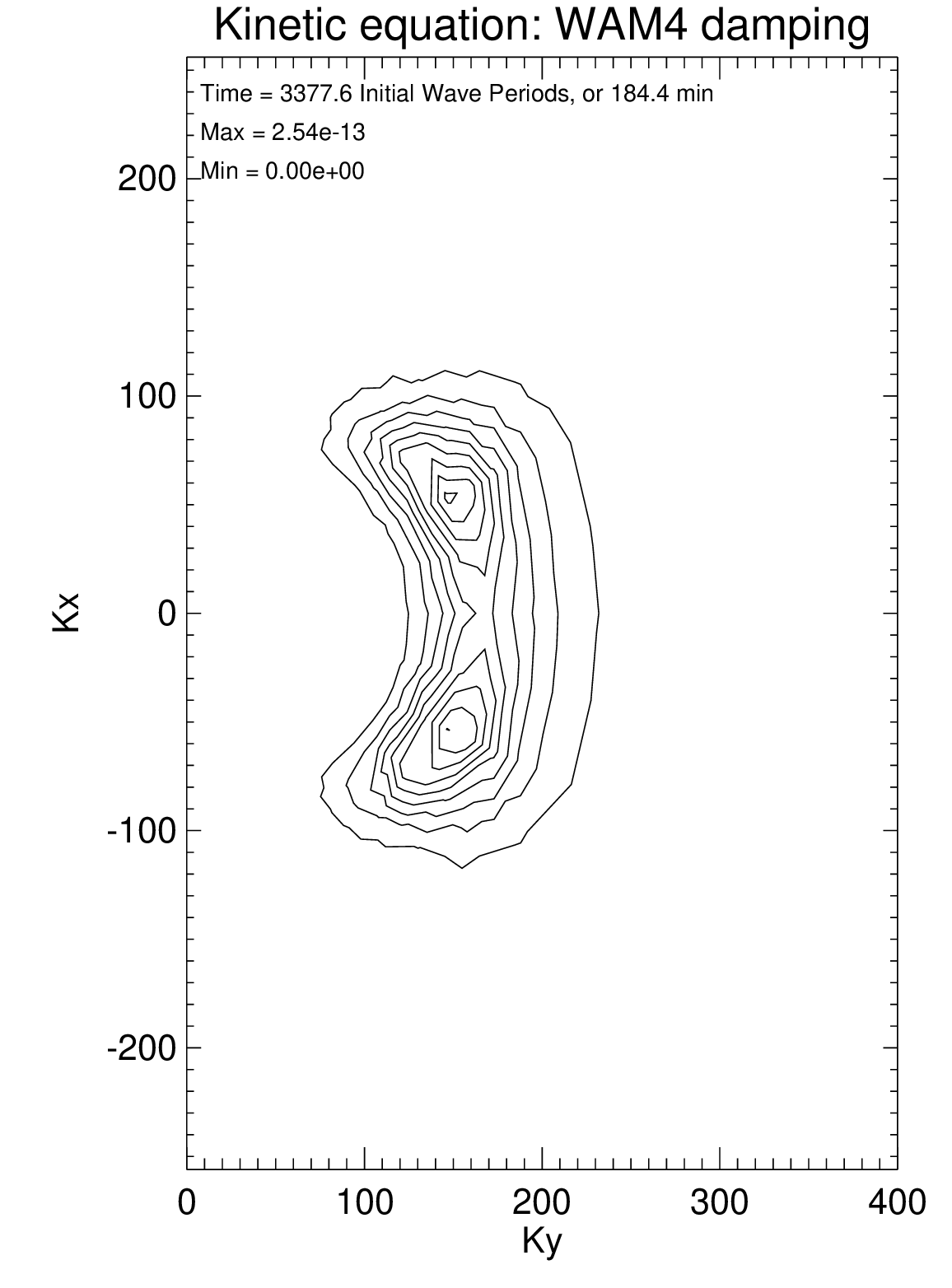}
\caption{(left)\label{KinEndWAM3}\label{KinEndWAM4}Level lines of the kinetic spectra in the {\it WAM3} (left) {\it WAM4} and damping cases at $t=3378.2\,T_0$.}
\end{figure}

Level lines of the low-pass filtered dynamic and kinetic spectrum at the beginning of the simulation are
presented on Fig.\ref{DynStart}-\ref{KinStart}. There are ten isolines on every figure. Values of level at
maximum and minimum isolines are shown on every picture.
Level lines of the low-pass filtered dynamic and kinetic spectrum for three different damping terms at the end of the simulation are presented on Fig.\ref{DynEnd}-\ref{KinEndWAM4}.
We observed development of bimodality in both experiments. This is in accordance with field
observations~\cite{Ewans1998}, \cite{WH2001}.

\begin{figure}[htb] 
\centering
\includegraphics[width=11.5cm,angle=0]{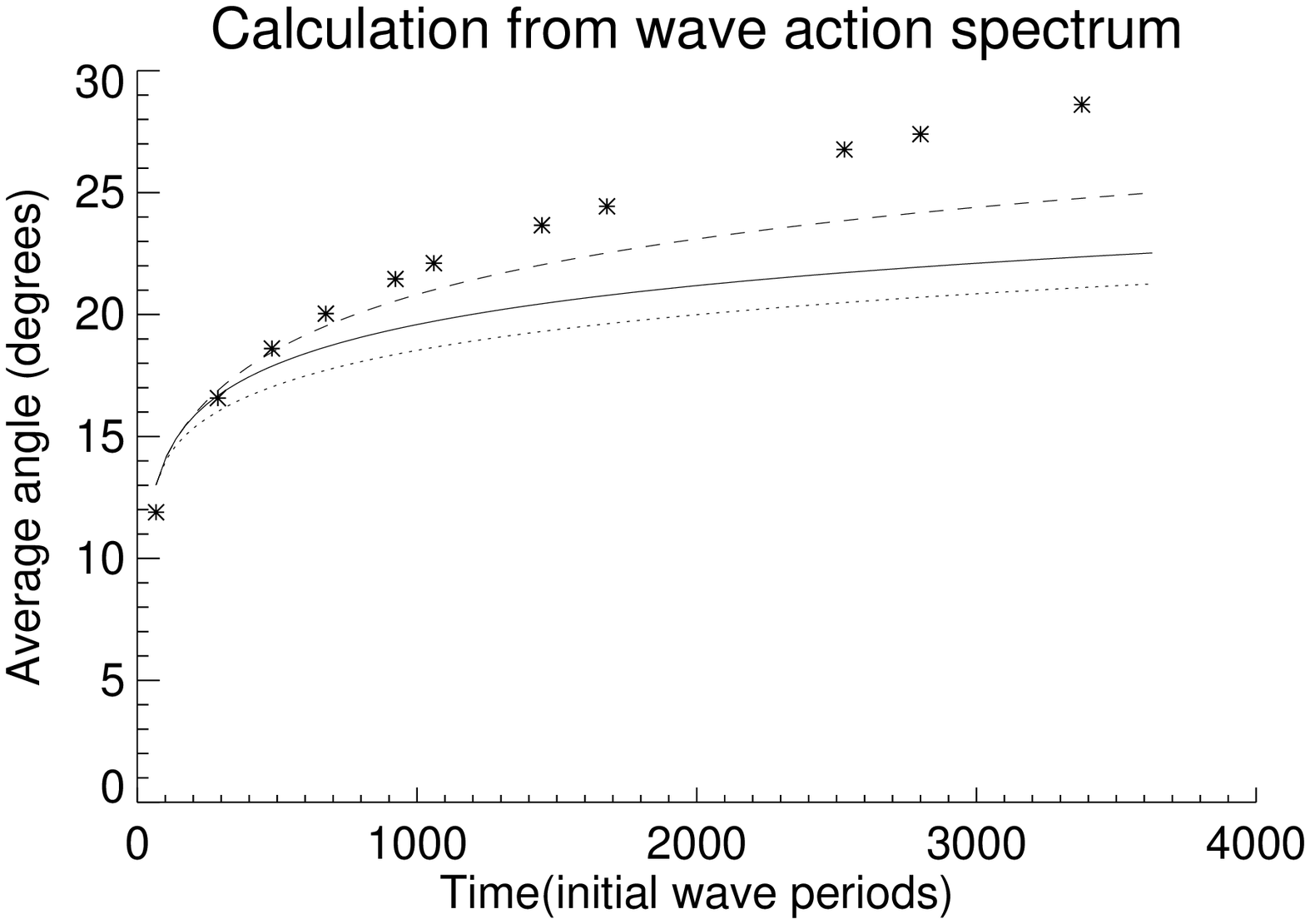} 
\caption{\label{SpreadCompareAction}Average angular spreading $\left(\int |\theta| n(\vec k) \D\vec{k}\right)/\left(\int n(\vec k) \D\vec{k}\right)$ as a function of time, calculated from wave action. Solid line - {\it WAM3}, dotted line - {\it WAM4}, dashed line - artificial viscosity, stars - dynamical equations}
\end{figure}

\begin{figure}[htb]
\centering
\includegraphics[width=11.5cm,angle=0]{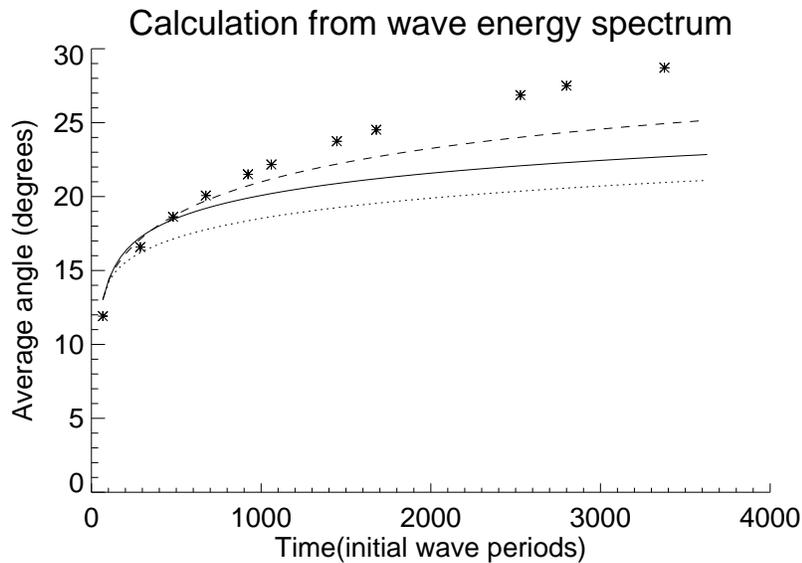} 
\caption{\label{SpreadCompareEnergy}
Average angular spreading $\left(\int |\theta| \omega n(\vec k) \D\vec{k}\right)/\left(\int \omega n(\vec k) \D\vec{k}\right)$ as a function of time, calculated from wave energy. Solid line - {\it WAM3}, dotted line - {\it WAM4}, dashed line - artificial viscosity, stars - dynamical equations}
\end{figure}

One can see good correspondence between the results of both experiments. Comparison
of time-evolution of the mean angular spreading, calculated from action and energy
spectra is presented on Fig. \ref{SpreadCompareAction}-\ref{SpreadCompareEnergy}.
\clearpage
We see growing divergence between dynamic and kinetic models. But using {\it WAM3} and {\it WAM4} models leads
to worse divergence. This is an additional argument against these variants of white-cap damping.

One can expect that the angular spreading will be arrested at later times, and the spectra will take universal self-similar shape.


\section{Conclusion} 
\begin{enumerate} 
\item{} Our numerical experiment shows that the Hasselmann equation without any additional terms is applicable
for description of gravity waves turbulence in the infinite basin only if nonlinearity is low enough. The average
wave steepness $\mu=({E^2}/{N^2})sqrt{2E}$ should be significantly less a certain critical level $\mu_0 = 0.095$.
To describe the wave turbulence of a moderate steepness $0.095 < \mu < 0.15$, one should add to the Hasselmann
equation some additional term $S_{diss}$ modeling dissipation due to white-capping. So far there is no any
theory making possible to find $S_{diss}$ analytically. This is a great challenge for hydrodynamicists. So far
we can suggest that $S_{diss}$ depends on steepness very sharply. Hence the guess that white-capping is the
threshold phenomenon formulated by Banner, Babanin, and Young~\cite{BBY2000} looks very plausible.
We did not model
too steep waves, but we can conjecture that if $\mu > 0.15$, white capping is the leading nonlinear process and
the weak-turbulent approach is not applicable at al.

\item{} Our results are not a surprise for designers of operational models for wave forecasting. They understood
necessity to introduce into the models extra dissipative terms many years ago. Another story that the models
of $S_{diss}$ routinely used in the WAM and WAVEWATCH models are too crude and do not grasp the most important
feature of the white-capping --- its threshold nature. In our opinion existing models of $S_{diss}$ overestimate
dissipation at small values of steepness and underestimate it at large $\mu$. Moreover, they overestimate
dissipation in the area of the spectral peak and underestimate it in the spectral region of high wave numbers.
We are planning to offer our own form of $S_{diss}$ extracted from our massive numerical simulation 
of Euler equation, but it will take some time and toil.

\item{} We have to stress that we solved maximally idealized model. We did not take into account interaction
of swell with atmosphere (see for instance~\cite{Drennan1999},~\cite{Grachev2003} ), interaction with ocean
currents, deviation of the wave motion from potentiality
as well as influence of turbulence in the surface-atmosphere boundary layer. In the future we plan to establish
a contact with experimentalists to develop more realistic model of swell propagation in the real ocean.

\item{} In one aspect our conclusions are very resolute. All existing experimental wave tanks cannot be
used for modelling of the waves propagation in open sea. The mesoscopic effects in wave tanks are too strong for
reasonable values of steepness. This pertains only to modeling of kinetics of energy containing region in the
vicinity of spectral peak. The rear faces of spectral distributions, spectral tails, can be successfully modelled.
This was demonstrated by Toba in his classical work~\cite{Toba1973}. Toba observed the Zakharov-Filonenko
spectrum $I_{\omega} \sim \omega^{-4}$ in the wave tank of moderate size. In our opinion this conclusion
is very important. Some authors, trying to find the universal law of fetch dependence of mean energy and mean
frequency put together data collected in ocean and in the experimental tanks. This mixed data hardly can be
compared with any self sufficient theory. This question is discussed in details in the article~\cite{BBZR2007}.

\end{enumerate}
Another conclusion is more pessimistic. The results of numerical experiments show that it is very difficult to reproduce real ocean conditions in laboratory wave tank. Even the size of the tanks of $200\times200$ meters is not large enough to model ocean due to the presence of wave numbers grid discreteness. 
 
\section{Acknowledgments} 
This work was supported by ONR grant N00014-06-C-0130, US Army Corps of Engineers Grant W912HZ-04-P-0172, RFBR grant 06-01-00665-a, INTAS grant 00-292, the Programme ``Nonlinear dynamics and solitons'' from the RAS Presidium and ``Leading Scientific Schools of Russia" grant, also by  and by NSF Grant NDMS0072803. We use this opportunity to greatfully acknowledge the support of these foundations.

A.\,O.~Korotkevich was also supported by Russian President grant for young scientist MK-1055.2005.2.
 
Also authors want to thank the creators of the open-source fast Fourier transform library FFTW~\cite{FFTW} for this fast, portable and completely free piece of software. 

\appendix

\section{\label{APP:ForbesList}``Forbes'' list of 15 largest harmonics.} 
Here one can find 15 largest harmonics at the end of calculations in the 
framework of dynamical equations. 
Average square of amplitudes in dissipation-less region was taken from smoothed 
spectrum, while all these harmonics exceed level $|a_{\vec k}|^2 = 1.4\times10^{-12}$.
\begin{tabular}{|c|c|c|c|c|}
\hline
$K_x$ & $K_y$ & $|a_{\vec k}|^2$ & $<|a_{\vec k}|^2>_f$ & $\frac{|a_{\vec k}|^2}{<|a_{\vec k}|^2>}$\\
\hline
-59 & 155 & 1.563e-12 & 0.746e-13 & 2.095e+1 \\
\hline
-37 & 166 & 1.903e-12 & 1.201e-13 & 1.585e+1 \\
\hline
-37 & 185 & 1.569e-12 & 2.288e-13 & 0.686e+1 \\
\hline
-36 & 162 & 1.477e-12 & 0.992e-13 & 1.489e+1 \\
\hline
-33 & 157 & 1.442e-12 & 0.713e-13 & 2.022e+1 \\
\hline
-26 & 164 & 3.351e-12 & 0.847e-13 & 3.956e+1 \\
\hline
-17 & 189 & 1.463e-12 & 2.789e-13 & 0.525e+1 \\
\hline
-14 & 173 & 1.408e-12 & 1.459e-13 & 0.965e+1 \\
\hline
-2 & 176 & 1.533e-12 & 1.697e-13 & 0.903e+1 \\
\hline
0 & 177 & 2.066e-12 & 1.741e-13 & 1.187e+1 \\
\hline
10 & 179 & 1.427e-12 & 1.893e-13 & 0.754e+1 \\
\hline
27 & 163 & 1.483e-12 & 0.832e-13 & 1.782e+1 \\
\hline
31 & 174 & 1.431e-12 & 1.342e-13 & 1.066e+1 \\
\hline
37 & 173 & 1.578e-12 & 1.581e-13 & 0.998e+1 \\
\hline
60 & 133 & 1.565e-12 & 0.345e-13 & 4.536e+1 \\
\hline
\end{tabular}

\section{\label{APP:FromDynamics2Hasselmann} From Dynamical Equations to Hasselmann Equation.} 
Standard setup for numerical simulation of the dynamical equations (\ref{eta_psi_equations}) implies $2 \pi \times 2 \pi$ domain in real space and gravity acceleration $g=1$. Usage of the domain size equal $2 \pi$ is convenient because in this case wave numbers are integers. 
 
In the contrary to dynamical equations, the kinetic equation (\ref{Hasselmann_equation}) is formulated in terms of real physical variables and it is necessary to describe the transformation from the ``dynamical'' variables into to the ``physical'' ones. 
 
Eq.\ref{eta_psi_equations} are invariant with respect to ``stretching'' 
transformation from ``dynamical'' to ``real'' variables: 
\begin{eqnarray} 
\eta_{\vec{r}} &=& \alpha \eta_{\vec{r}^\prime}^\prime,\,\,\,\,\vec{k} = 
\frac{1}{\alpha} \vec{k}^\prime,\,\,\,\,\vec{r} = \alpha 
\vec{r}^\prime,\,\,\,\,g = \nu g^\prime, \\ 
t &=& \sqrt{\frac{\alpha}{\nu}}t^\prime,\,\,\, L_x = \alpha L_x^\prime,\,\,\, 
L_y = \alpha L_y^\prime 
\end{eqnarray} 
where prime denotes variables corresponding to dynamical equations. 
 
In current simulation we used the stretching coefficient $\alpha=800.00$, which allows to reformulate the statement of the problem in terms of real physics: we considered $5026\,m \times 5026\,m$ periodic boundary conditions domain of statistically uniform ocean with the same resolution in both directions and characteristic wave length of the initial condition around $22\, m$. In oceanographic terms, this statement corresponds to the ``duration-limited experiment''.

\end{document}